\documentclass[12pt]{iopart}
\usepackage{iopams}                                      
\usepackage{mathrsfs}                                    
\usepackage[latin9]{inputenc}                            
\usepackage[english]{babel}                              
\usepackage[T1]{fontenc}                                 
\usepackage{setspace}                                    
\usepackage{enumitem}                                    
\usepackage{color}                                       
\usepackage{graphicx}                                    
\usepackage{subfigure}                                   
\usepackage{cite}                                        
\usepackage{multirow}                                    
\usepackage{scalerel,stackengine}                        

\newcommand{\tn}{\textnormal}

\def\jnl@style{\it}
\def\aaref@jnl#1{{\jnl@style#1}}

\def\aaref@jnl#1{{\jnl@style#1}}

\def\aj{\aaref@jnl{AJ}}                   
\def\araa{\aaref@jnl{ARA\&A}}             
\def\apj{\aaref@jnl{ApJ}}                 
\def\apjl{\aaref@jnl{ApJ}}                
\def\apjs{\aaref@jnl{ApJS}}               
\def\ao{\aaref@jnl{Appl.~Opt.}}           
\def\apss{\aaref@jnl{Ap\&SS}}             
\def\aap{\aaref@jnl{A\&A}}                
\def\aapr{\aaref@jnl{A\&A~Rev.}}          
\def\aaps{\aaref@jnl{A\&AS}}              
\def\azh{\aaref@jnl{AZh}}                 
\def\baas{\aaref@jnl{BAAS}}               
\def\jrasc{\aaref@jnl{JRASC}}             
\def\memras{\aaref@jnl{MmRAS}}            
\def\mnras{\aaref@jnl{MNRAS}}             
\def\pra{\aaref@jnl{Phys.~Rev.~A}}        
\def\prb{\aaref@jnl{Phys.~Rev.~B}}        
\def\prc{\aaref@jnl{Phys.~Rev.~C}}        
\def\prd{\aaref@jnl{Phys.~Rev.~D}}        
\def\pre{\aaref@jnl{Phys.~Rev.~E}}        
\def\prl{\aaref@jnl{Phys.~Rev.~Lett.}}    
\def\pasp{\aaref@jnl{PASP}}               
\def\pasj{\aaref@jnl{PASJ}}               
\def\qjras{\aaref@jnl{QJRAS}}             
\def\skytel{\aaref@jnl{S\&T}}             
\def\solphys{\aaref@jnl{Sol.~Phys.}}      
\def\sovast{\aaref@jnl{Soviet~Ast.}}      
\def\ssr{\aaref@jnl{Space~Sci.~Rev.}}     
\def\zap{\aaref@jnl{ZAp}}                 
\def\nat{\aaref@jnl{Nature}}              
\def\iaucirc{\aaref@jnl{IAU~Circ.}}       
\def\aplett{\aaref@jnl{Astrophys.~Lett.}} 
\def\apspr{\aaref@jnl{Astrophys.~Space~Phys.~Res.}}
\def\bain{\aaref@jnl{Bull.~Astron.~Inst.~Netherlands}} 
\def\fcp{\aaref@jnl{Fund.~Cosmic~Phys.}}  
\def\gca{\aaref@jnl{Geochim.~Cosmochim.~Acta}}   
\def\grl{\aaref@jnl{Geophys.~Res.~Lett.}} 
\def\jcp{\aaref@jnl{J.~Chem.~Phys.}}      
\def\jgr{\aaref@jnl{J.~Geophys.~Res.}}    
\def\jqsrt{\aaref@jnl{J.~Quant.~Spec.~Radiat.~Transf.}}
\def\memsai{\aaref@jnl{Mem.~Soc.~Astron.~Italiana}}
\def\nphysa{\aaref@jnl{Nucl.~Phys.~A}}   
\def\physrep{\aaref@jnl{Phys.~Rep.}}   
\def\physscr{\aaref@jnl{Phys.~Scr}}   
\def\planss{\aaref@jnl{Planet.~Space~Sci.}}   
\def\procspie{\aaref@jnl{Proc.~SPIE}}   

\graphicspath{{Figures/}}

\newcommand\equalhat{\mathrel{\stackon[1.5pt]{=}{\stretchto{\scalerel*[\widthof{=}]{\wedge}{\rule{1ex}{3ex}}}{0.5ex}}}}
\newcommand{\RN}[1]{\textup{\uppercase\expandafter{\romannumeral#1}}}

\begin{document}

\title{Gravitational geons in asymptotically anti-de Sitter spacetimes}
\author{Gr\'egoire Martinon$^1$}
\author{Gyula Fodor$^2$}
\author{Philippe Grandcl\'ement$^1$}
\author{Peter Forg\'acs$^{2,3}$}

\address{$^1$ LUTH, Observatoire de Paris, PSL Research University, CNRS, Universit\'e Paris Diderot, Sorbonne Paris Cit\'e, 92190 Meudon, France}
\address{$^2$ Wigner Research Centre for Physics, RMKI, 1525 Budapest 114, P.O. Box 49, Hungary}
\address{$^3$ LMPT, CNRS-UMR 6083, Universit\'e de Tours, Parc de Grandmont, 37200 Tours, France}

\ead{gregoire.martinon@obspm.fr}

\date{\today}

\begin{abstract}
We report on numerical constructions of fully non-linear geons in asymptotically anti-de Sitter (AdS) spacetimes in four
dimensions. Our approach is based on 3+1 formalism and spectral methods in a gauge combining maximal slicing and spatial harmonic
coordinates. We are able to construct several families of geons seeded by different families of spherical harmonics. We can reach
unprecedentedly high amplitudes, with mass of order $\sim 1/2$ of the AdS length, and with deviations of the order of 50\%
compared to third order perturbative approaches. The consistency of our results with numerical resolution is carefully checked and
we give extensive precision monitoring techniques.  All global quantities like mass and angular momentum are computed using two
independent frameworks that agree each other at the $0.1\%$ level. We also provide strong evidence for the existence of excited
(i.e. with one radial node) geon solutions of Einstein equations in asymptotically AdS spacetimes by constructing them
numerically.
\end{abstract}

\submitto{\CQG}

\maketitle

\section{Introduction}
\label{intro}

AdS spacetime has drawn a great deal of attention since the emergence of gauge/gravity duality \cite{Maldacena98, Witten98,
Aharony00, Hubeny15}, which basically states that string theory in asymptotically AdS spacetimes is equivalent to a conformal
field theory (CFT) living on the AdS timelike boundary. The non-linear stability of the AdS spacetime is far from being settled
and is of great interest.

In the seminal paper \cite{Bizon11}, the time-evolution of a free, massless scalar field coupled to
Einstein's gravity has been investigated in asymptotically AdS spacetimes. The authors found that for a large set of smooth
initial data black holes form, indicating that AdS is unstable against black hole formation. It has been conjectured that no
matter how small perturbation one considers in asymptotically AdS, it will eventually lead to the formation of a black hole.
Various analytical studies and numerical experiments followed
\cite{Jalmuzna11,Buchel12,Maliborski12,Bizon14,Maliborski14,Friedrich14,Bizon15,Buchel15,Deppe15a,Craps14,
Craps15,Evnin15,Deppe16,Gursoy16,Olivan16a,Olivan16b,Brito16,Menon16,Evnin16,Dimitrakopoulos17a,Dimitrakopoulos17b}, accumulating
more and more indications to the conjectured instability. Since asymptotically AdS spacetimes have timelike boundaries, by
imposing energy-conserving boundary conditions, outgoing waves will bounce back from spatial infinity and reach the origin in
finite time. The heuristic explanation for the instability of asymptotically AdS spacetimes is that by bouncing from the boundary
many times, the waves concentrate more and more energy to smaller and smaller scales, leading to the formation of a black hole in
a timescale proportional to the inverse square of the initial amplitude.

In a number of papers \cite{Maliborski13,Fodor14,Fodor15} it has been found that a real massless scalar field minimally coupled to
gravity admits asymptotically AdS globally regular, spherically symmetric, time-periodic solutions, whose mass is finite. The existence of
families of localized, time-periodic asymptotically AdS solutions (referred to as scalar AdS breathers) is intimately linked to the boundary
conditions. In asymptotically flat spacetimes, breathers represent the exception to the rule ``anything that
can radiate, does radiate'', and they exist only in rather special circumstances (e.g. sine-Gordon theory in 1+1 dimension,
scalar theories with V-shaped \cite{Arodz06,Arodz08} or logarithmic potentials \cite{Koutvitsky06}). On the other hand,
families of long-lived, closely time-periodic {\sl oscillons} exist in a number of field theories containing massive scalars
in Minkowski or asymptotically flat spacetimes \cite{Bogolyubskii77,Gleiser94,Copeland95}. Oscillons emit energy very slowly,
accompanied by a slow change of their amplitude and frequency \cite{Honda02,Fodor06,Fodor08}. It has been shown that the energy
emission rate of small-amplitude oscillons is exponentially small in terms of a parameter corresponding to the central amplitude
\cite{Segur87,Fodor09a,Fodor09b}. Flat background oscillons only exist for massive fields, since their frequency at small amplitude
is determined by the scalar field mass. Massive scalar fields coupled to Einstein's gravity also form long living,
oscillating localized objects, referred to as oscillatons \cite{Seidel91,Seidel94,Page04}. The mass loss rate of oscillatons can
be extremely small even on cosmological time scales \cite{Fodor10,Grandclement11}. Importantly oscillons/oscillatons form from
generic initial data and are stable under time evolution.

Many years ago J. A. Wheeler introduced the concept of {\sl geon}, a hypothetical, time-dependent solution of either Einstein or
Einstein-Maxwell equations. The gravitational geon would consist of high frequency gravitational
waves, trapped in a background geometry created by the waves themselves {\sl for much longer times} than the light crossing time
\cite{Wheeler55}. We refer to \cite{Brill64,Anderson97} for approximation proceedures of gravitational geon in
asymptotically flat space-times using frequency averaging and self-consistent methods. A theorem by G.W. Gibbons and J.M. Stuart
\cite{Gibbons84} has established the absence of asymptotically flat solutions of Einstein equations which are time-periodic and
empty near infinity, implying that if geons exist, they cannot be time-periodic i.e. they should necessarily radiate. This does
not exclude the possibility that asymptotically flat geons would loose their mass slowly similarly to what has been found for
oscillons/oscillatons.

The simplest and mostly investigated scalar AdS breathers are spherically symmetric and importantly some of them appear to be
actually stable against collapsing to a black hole
\cite{Buchel13,Maliborski13,Green15,Fodor14,Deppe15b,Fodor15,Duarte16,Herdeiro16}. A key ingredient for the construction of such
solutions seems to be a large concentration of the energy into one single scalar eigenmode. The existence of such stable breathers
indicate the presence of stability islands in asymptotically AdS spacetimes, likely forming sets of non-zero measure initial data.

In \cite{Dias12a}, it has been shown by a series expansion of the metric tensor that time-periodic solutions of Einstein's
equation with a negative cosmological constant, ``gravitational AdS breathers'', or AdS {\sl geons} are likely to exist. The
perturbative expansion becomes inconsistent at third order, with the outbreak of a secular resonance. However, a standard
Poincar\'e-Lindstedt method, which consist in promoting the geon frequency to a function of the amplitude, can be applied to cure
this inconsistency, at least in somes cases \cite{Dias16a,Rostworowski16,Rostworowski17}. This suggests that fully non-linear geons can be
constructed, and numerical solutions of the simplest family of geons were constructed in \cite{Horowitz14}. Geons can be seen as
non-linear purely gravitational eigenmodes of asymptotically AdS spacetimes, being the fundamental excitations resisting the
collapse to a black hole. They are among the key ingredients to better understand the AdS instability problem.

The main results of the present paper are the followings. (i) we give an independent construction of the fully non-linear $(l,m,n) = (2,2,0)$ geons
that were constructed solely in \cite{Horowitz14}. Our results on global quantities disagree, to a certain extent, with this work,
but strongly supported by convergent analytical and numerical arguments(section \ref{res}). (ii) we present the so-called
Andersson-Moncrief gauge and discuss its theoretical motivations as well as its numerical implementation. We also make clear the
link between this gauge and the harmonic gauge enforced with the popular De-Turck method (section \ref{gauge} and appendix
\ref{DeTurck}). (iii) we extend the numerical constructions of fully non-linear geons to the $(l,m,n) = (4,4,0)$ case, as well as
to the three excited families exhibiting one radial node. The existence of these excited geons was denied in \cite{Dias16a}, but
supported in \cite{Rostworowski16,Rostworowski17} with perturbative arguments. Our results clearly provide evidence in
favour of the existence of such solutions.

This paper is organised as follows. We carry out first a perturbative expansion around the AdS metric to construct approximations to AdS geons.
Our perturbative approach is based on the Kodama-Ishibashi formalism \cite{Kodama00,Kodama03,Kodama04,Ishibashi04}, which method
also has been used in \cite{Dias12a}. Several AdS geon families are constructed, corresponding to the quantum numbers $(l,m,n) =
(2,2,0)$ and $(4,4,0)$ as well as all three radially excited families with $m = 2$. Then we present in detail our numerical
construction of several different geon families, corresponding to different angular and radial excitations. As an initial guess,
we use perturbative results, summarized in \Sref{pert}. In \Sref{gauge} we present a novel approach inspired by Andersson and
Moncrief \cite{Andersson03} and based on a 3+1 decomposition of Einstein equations \cite{Gourgoulhon07} in a gauge combining
maximal slicing and spacial harmonic coordinates. \Sref{reg} is dedicated to our regularization procedure, as all fields in AdS
are diverging. \Sref{aads} is devoted to the definition of AdS asymptotics. In \Sref{numproc}, we describe the numerical
algorithm based on the Kadath spectral library \cite{Grandclement10}. In order to check the validity of our solutions, we build
up an efficient precision monitoring. We then develop several diagnostics to check the precision of our code, in particular we
ensure that our Dirichlet boundary conditions preserve the AdS asymptotics, in the sense of \cite{Ashtekar84,Ashtekar00}. Finally
in \Sref{res}, we build numerical geons at unprecedentedly high amplitudes, with deviations as high as 50\% relative to third
order perturbative approaches. We carefully extract global variables with two different methods, namely the Ashtekar-Magnon-Das
(AMD) \cite{Ashtekar84,Ashtekar00} and the Balasubramanian-Kraus (BK) \cite{Balasubramanian99} that agree each other at the
$0.1\%$ level. We recover in the low amplitude limit the perturbative results up to fifth order, and give predictions about
coefficients appearing at successive orders. Last but not least, we provide numerical evidence that radially excited geons, whose
existence was debated in \cite{Dias16a,Rostworowski16,Rostworowski17}, do exist in AdS spacetimes. We compute them numerically and
finally show their existence curves in phase space.

\section{Perturbative approach}
\label{pert}

We denote by $\bar{g}$ the 4-dimensional AdS metric. Following \cite{Kodama00,Kodama03,Kodama04,Ishibashi04}, we introduce coordinates that make
the rotational invariance of the background explicit, namely
\begin{eqnarray}
   d\bar s^2 &=& \bar{g}_{\alpha\beta}dx^\alpha dx^\beta = \mathring{g}_{ab}(y)dy^ady^b + r^2(y)\tilde{\gamma}_{ij}(z)dz^idz^j\\
             &=& \frac{L^2}{\cos^2 \bar x} \left[-d\bar t^2 + d\bar x^2 + \sin^2 \bar x (d\theta^2 + \sin^2\theta d\varphi^2)\right] ,
\end{eqnarray}
where $L$ is related to the cosmological constant $\Lambda$ by $\Lambda = -3/L^2$, $y^a$ coordinates span the
time-radial plane and $z^i$ the unit sphere $\mathcal{S}^2$ whose metric is $\tilde\gamma_{ij}$. By identification, we will
denote in this section $r = L \tan \bar x$. We look for families of solutions of the vacuum Einstein equation with a negative
cosmological constant in the following form:
\begin{equation}
 g_{\alpha\beta}=\bar g_{\alpha\beta}+\sum_{k=1}^\infty \varepsilon^k h_{\alpha\beta}^{(k)} ,
\end{equation}
where $\varepsilon$ is a small expansion parameter and $k$ denotes the orders of approximation. We assume that at infinity
\numparts
\begin{eqnarray}
   \lim_{\bar x\to\frac{\pi}{2}}\left[h^{(k)}_{\bar t\bar t}\left(\frac{\pi}{2}-\bar x\right)^2\right]&=&-L^2\nu_k, \label{nukdefeq}\\
   \lim_{\bar x\to\frac{\pi}{2}}\left[h^{(k)}_{\alpha\beta}\left(\frac{\pi}{2}-\bar x\right)^2\right]&=&0 \quad \tn{for } \alpha\neq \bar t \  \mathrm{or} \ \beta \neq \bar t,
\end{eqnarray}
\endnumparts
where $\nu_k$ are constants, independent of the angular coordinates, and they are generally nonzero only for even $k$. Then the
leading order behavior of the $g_{\bar t\bar t}$ metric component is $\varepsilon$ dependent,
\begin{equation}
 \lim_{\bar x\to\frac{\pi}{2}}\left[g_{\bar t\bar t}\left(\frac{\pi}{2}-\bar x\right)^2\right]=-\nu, \quad \nu=L^2\left(1+\sum_{k=1}^\infty\varepsilon^k\nu_k\right) . \label{eqnudef}
\end{equation}
The advantage of this setting is that (i) it maintains the AdS asymptotics (in the sense of \cite{Ishibashi04, Ashtekar00,
Ashtekar84} and \Sref{aads} below) and (ii) the frequency $\bar\omega$ of the geon can be kept independent of $\varepsilon$ at all
orders with respect to the time coordinate $\bar t$. The asymptotically AdS time coordinate is then $t=\bar t \sqrt{\nu}$, and
the physical frequency of the solution is $\omega=\bar\omega/\sqrt{\nu}$.

At each order we decompose the metric perturbation $h_{\alpha\beta}^{(k)}$ into the sum of scalar and vector spherical harmonic
components\footnote{Tensor spherical harmonics appear only in spacetimes of dimension at least five.}, in the same way as it has
been done for the linear case in \cite{Kodama00,Ishibashi04}. We choose a gauge in which scalar-type perturbations in the
$\mathbb{S}_{lm}$ real spherical harmonic class have the block diagonal form ($a,b=\bar t,\bar x$ and $i,j=\theta,\varphi$)
\begin{equation}
   h^{(k)}_{ab}=H^{(s)}_{ab}\mathbb{S} , \quad
   h^{(k)}_{ai}=0 , \quad
   h^{(k)}_{ij}=H^{(s)}_L\tilde\gamma_{ij}\mathbb{S}, \label{scalmetrpert}
\end{equation}
and class $\mathbb{V}_{(lm)i}$ vector-type perturbations are
\begin{equation}
h^{(k)}_{ab}=0 , \quad h^{(k)}_{ai}=H^{(v)}_{a}\mathbb{V}_i ,
\quad h^{(k)}_{ij}=0, \label{vectmetrpert}
\end{equation}
where we made implicit the $(l,m)$ indices and the order $k$ of the perturbation. Here the functions $H^{(s)}_{ab}$, $H^{(s)}_L$
and $H^{(v)}_{a}$ depend only on the coordinates $y^a=(\bar t,\bar x)$.
We use real spherical harmonics $\mathbb{S}_{lm}$ that are orthonormal. The $\phi$ dependence of $\mathbb{S}_{lm}$ for $m\geq0$ is
$\cos(m\phi)$, and for $m<0$ it is $\sin(|m|\phi)$. The vector harmonics $\mathbb{V}_{(lm)i}$ can be expressed in terms of the
scalar harmonics as
\begin{equation}
 \mathbb{V}_{(lm)\theta}=\frac{1}{\sqrt{l(l+1)}}
 \frac{1}{\sin\theta}\frac{\partial \mathbb{S}_{lm}}{\partial\phi} \ , \quad
 \mathbb{V}_{(lm)\phi}=\frac{-1}{\sqrt{l(l+1)}}
 \sin\theta\frac{\partial\mathbb{S}_{lm}}{\partial\theta} .
\end{equation}

Perturbative modes with $l\leq 1$ have to be treated separately. For the $l=0$ scalar modes we further restrict the gauge by assuming
$H^{(s)}_L=0$ and $H^{(s)}_{\bar t\bar x}=0$. For the $l=1$ vector mode we set $H^{(v)}_{\bar x}=0$.

For $l\geq 2$, the Kodama-Ishibashi-Seto\cite{Kodama00} gauge-invariant variables $Z$, $Z_a$ and $Z_{ab}$ can be defined, and in our
gauge they are related to the metric variables as
\begin{equation}
   H^{(s)}_L=\frac{r^2}{2}Z , \quad
   H^{(s)}_{ab}=Z_{ab}-\frac{1}{2}Z \mathring g_{ab} , \quad
   H^{(v)}_{a}=Z_a  \ .
   \label{eqshhzz}
\end{equation}
In each scalar and vector class, perturbations are characterized by functions $\Phi^{(s)}$ and $\Phi^{(v)}$ respectively,
satisfying the following master equation, deduced from the Einstein equation:
\begin{equation}
   -\partial^2_{\bar t}\Phi^{(s,v)}+\partial^2_i{\bar x}\Phi^{(s,v)} -\frac{l(l+1)}{\sin^2 \bar x}\Phi^{(s,v)}+\frac{\Phi^{(s,v)(0)}}{\sin^2 \bar x}=0,
   \label{eqmasterphi}
\end{equation}
where $\Phi^{(s,v)(0)}$ are inhomogeneous source terms fixed by lower order perturbations. The outer boundary conditions for the
scalar and vector cases are respectively
\begin{equation}
   \lim_{\bar x\to\frac{\pi}{2}} \partial_{\bar x} \Phi^{(s)} = 0 \quad \tn{and} \quad \lim_{\bar x\to\frac{\pi}{2}}\Phi^{(v)} = 0 .
\end{equation}

From the scalar-type generating function, defining $\phi^{(s)} = r\Phi^{(s)}$, we obtain the gauge-invariant variables as
\numparts
\begin{eqnarray}
   Z_{\bar t\bar t}&=&\partial^2_{\bar t}\phi^{(s)}-\tan\bar x\partial_{\bar x}\phi^{(s)} +\frac{\phi^{(s)}}{\cos^2\bar x}+Z_{\bar t\bar t}^{(0)}, \label{zscalareqs1}\\
   Z_{\bar t\bar x}&=&\partial_{\bar t}\partial_{\bar x}\phi^{(s)}
   -\tan\bar x\partial_{\bar t}\phi^{(s)}+Z_{\bar t\bar x}^{(0)},\\
   Z_{\bar x\bar x}&=&\partial^2_{\bar x}\phi^{(s)}-\tan\bar x\partial_{\bar x}\phi^{(s)} -\frac{\phi^{(s)}}{\cos^2\bar x}+Z_{\bar x\bar x}^{(0)}, \\
   Z&=&\frac{\cos^2 \bar x}{L^2}\left(Z_{\bar x\bar x}-Z_{\bar t\bar t}\right)+Z^{(0)}\ , \label{zscalareqs4}
\end{eqnarray}
\endnumparts
where the inhomogeneous source terms $Z_{ab}^{(0)}$ and $Z^{(0)}$ can be computed from lower order perturbation results. The
vector-type gauge-invariant variable can be obtained from the vector-type generating function $\phi^{(v)} = r\Phi^{(v)}$ as
\begin{equation}
   Z_{\bar t}=\partial_{\bar x}\phi^{(v)}+Z_{\bar t}^{(0)} , \quad
   Z_{\bar x}=\partial_{\bar t}\phi^{(v)}\ , \label{zvectoreqs}
\end{equation}
where $Z_{\bar t}^{(0)}$ is an other inhomogeneous source term. Finally, the metric perturbation variables in each class can be
recovered with \Eref{eqshhzz}.

At first order in the $\varepsilon$ expansion no source terms appear from lower orders, and the scalar equation can be solved explicitly.
Regular and asymptotically AdS scalar-type perturbations exist only for frequencies $\bar\omega_s=l+1+2n$, with $n\geq 0$
integer. The corresponding generating function is
\begin{equation}
   \Phi^{(s)}_1=\frac{\alpha}{L}\sin^{l+1}\bar x\,\cos(\bar\omega_s \bar t-\delta) \frac{n!}{(l+\frac{3}{2})_{n}} P_{n}^{(l+\frac{1}{2},-\frac{1}{2})}(\cos(2\bar x)),
   \label{eqgenfscal}
\end{equation}
where $\alpha$ is a constant figuring the amplitude, $\delta$ is a constant phase, $P$ are the Jacobi polynomials, and the
Pochhammer's Symbol is $(c)_n=\Gamma(c+n)/\Gamma(c)$. Vector-type perturbations exist for frequencies $\bar\omega_v=l+2+2n$,
having the form
\begin{equation}
   \Phi^{(v)}_1=\frac{\alpha}{L}\cos\bar x\sin^{l+1}\bar x\,\cos(\bar\omega_v \bar t-\delta) \frac{n!}{(l+\frac{3}{2})_{n}} P_{n}^{(l+\frac{1}{2},\frac{1}{2})}(\cos(2\bar x)).
   \label{eqgenfvect}
\end{equation}
Hereafter, the integer $n$ will denote the radial excitation number.

As a valid first order solution, we can choose any linear combination of the metric perturbations generated by equations \eref{eqgenfscal} and
\eref{eqgenfvect}, with arbitrary amplitudes and phases, for a number of different $(l,m)$. Proceeding to second order in
$\varepsilon$, there will be $\Phi^{(s,v)(0)}$ source terms appearing in \Eref{eqmasterphi} for certain scalar and vector $(l,m)$
cases. At $\varepsilon^2$ order these equations are generally solvable. However, at third order,  certain $\Phi^{(s,v)(0)}$ will
contain terms that have time dependence with resonant frequencies $\bar\omega_{s,v}$. In these resonant cases, the master
\Eref{eqmasterphi} can have regular asymptotically AdS time-periodic solutions only if certain consistency conditions hold (they
generally require the vanishing of an integral between $\bar x=0$ and $\bar x=\pi/2$ involving the source term). These make severe
restrictions on the allowed amplitudes and phases of the linear modes that we include at first order in $\varepsilon$.
Furthermore, in these resonant cases the solution for $\Phi^{(s,v)}$ is not unique, one can always add a constant times the
homogeneous solution. This implies that at $\varepsilon^3$ order new constants appear in the expansion procedure, and these
constants will be restricted later by the consistency conditions at order $\varepsilon^5$.

It is natural to start with as few components as possible at the linear level.
On the other hand, it is important to see what kind of combinations of same-frequency linear modes can survive to higher order in the formalism.
For example, one can try to find the most general
monochromatic scalar-type solution with $l=2$ and $n=0$ that has a mirror symmetry at the $\theta=\pi/2$ plane. At the linear level,
any combination of the $m=2,-2,0$ modes with the independent $\delta = 0$ and $\pi/2$ phases is allowed (i.e. six independent
amplitudes). Solving the consistency conditions at $\varepsilon^3$ order, it turns out that all solutions in this class with a nonzero angular
momentum $J$ must be helically symmetric.

In the present paper we focus on helically symmetric geons. In particular, we study in more details the one-parameter family
of helically symmetric geons arising from the $l=2$, $m=\pm2$, $n=0$ linear scalar modes, which was already considered in
\cite{Dias12a} and \cite{Horowitz14}. In order to build helically symmetric perturbative geons, we combine the metric
perturbations corresponding to $(m = 2,\delta = 0)$ and $(m=-2,\delta = \pi/2)$ with equal amplitude, such that the nonzero
components of the metric perturbation are
\begin{eqnarray}
 h_{\bar t\bar t}&=&h_{\bar x\bar x}=-9L^2\frac{\sin^2\bar x\cos(2\bar x)}{2\cos\bar x}
  \sin^2\theta\cos(3\bar t-2\phi) \ , \\
 h_{\bar t\bar x}&=&9L^2\sin^3\bar x\sin^2\theta\sin(3\bar t-2\phi) \ , \\
 h_{\theta\theta}&=&\frac{h_{\phi\phi}}{\sin^2\theta}=-9L^2\frac{\sin^4\bar x}{2\cos\bar x}
  \sin^2\theta\cos(3\bar t-2\phi) \ .
\end{eqnarray}
A simple change of coordinates $\phi \leftarrow \phi - \omega \bar t/m$ brings the metric in a manifestly time independent form
whose Killing vector is

\begin{equation}
   \partial_{t'} \equiv \partial_{\bar t} + \frac{\omega}{m}\partial_{\varphi}.
\end{equation}

We were able to unambiguously construct the $\varepsilon$ expansion of the metric for this geon up to fourth order. In order to
achieve this we had to solve the consistency conditions at fifth order in $\varepsilon$, which is necessary for getting the
concrete values of the integration constants that appear at third order in the expansion. For rotating geons a natural way to fix
the re-parametrization freedom in the $\varepsilon$ parameter is to cancel all higher order coefficients in the expansion of the
angular momentum $J$. Choosing then the coefficient $\alpha$ in \Eref{eqgenfscal} appropriately, we set $J = 27\pi
L^2\varepsilon^2/128$, which agrees with the choice made in \cite{Dias12a}. The obtained expansions for the frequency and the mass
(see \Sref{aads} below for definition) of the configuration can be written as
\numparts
\begin{eqnarray}
   \omega L &=&3\left(1+\omega_1 \frac{J}{L^2} + \omega_2 \frac{J^2}{L^4} + \ldots\right),\\
   \label{omjexpansion}
   \omega L &=&3\left(1+\omega_1 \frac{2M}{3L} + \frac{2}{3} (2\omega_2-\omega_1^2)\frac{M^2}{L^2} + \ldots\right),\\
   \frac{M}{L} &=&\frac{3}{2}\left(\frac{J}{L^2} + \frac{\omega_1}{2} \frac{J^2}{L^4} + \frac{\omega_2}{3} \frac{J^3}{L^6}+\ldots\right) ,
   \label{mjexpansion}
\end{eqnarray}
\endnumparts
where
\numparts
\begin{eqnarray}
   \omega_1&=&-\frac{4901}{3780\pi}\approx-0.412708,\\
   \omega_2&=&\frac{7823862709549425\pi^2-76880912765261056} {73229764608000\pi^2}\approx0.466991.
\end{eqnarray}
\endnumparts
In order to get the $J^3$ term in the mass we had to calculate the $l=0$ and $l=1$ components at sixth order in $\varepsilon$.

In the remaining of the paper, first order geons are used as initial guess configurations. Successive perturbative orders allow us
to quantify the proximity between the analytical and numerical approaches.

\section{Gauge-fixing}
\label{gauge}

\subsection{Pertubative geon in the Andersson-Moncrief gauge}

We start with the 3+1 formalism \cite{Gourgoulhon07}, where the metric reads
\begin{equation}
   g_{\alpha\beta}dx^\alpha dx^\beta = -N^2dt^2 + \gamma_{ij}(dx^i + \beta^idt)(dx^j + \beta^jdt),
\end{equation}
where latin indices denote spatial directions on $t = cst$ hypersurfaces $\Sigma_t$, $N$ is the so-called lapse function,
$\beta^i$ the shift vector and $\gamma_{ij}$ the spatial 3-metric. The normal vector to $\Sigma_t$ is $u_\alpha = (-N,0,0,0)$. We
also introduce $K_{ij}$ the extrinsic curvature of $\Sigma_t$
\begin{equation}
   K_{ij} = -\frac{1}{2N}\mathcal{L}_m\gamma_{ij},
   \label{defK}
\end{equation}
where $\mathcal{L}_m$ is the Lie derivative in the direction $m^\alpha \equiv Nu^\alpha = \partial_t^\alpha - \beta^\alpha$.

In order to solve the system, one needs to specify appropriate gauge conditions. The foliation is defined by the maximal slicing
condition and the spatial coordinates are chosen to be harmonic. We dubbed this choice the Andersson-Moncrief gauge in
reference to \cite{Andersson03}. The Andersson-Moncrief gauge corresponds to imposing the following conditions:
\numparts
\begin{eqnarray}
   \label{K}
   K &\equiv& \gamma^{ij}K_{ij} = 0,\\
   V^i &\equiv& \gamma^{kl}(\Gamma_{kl}^i - \bar{\Gamma}_{kl}^i) = -\sqrt{\frac{\bar{\gamma}}{\gamma}} \bar{D}_k\left( \sqrt{\frac{\gamma}{\bar{\gamma}}}\gamma^{ik} \right) = 0,
   \label{V}
\end{eqnarray}
\endnumparts
where $\Gamma_{kl}^i$ is the Christoffel symbol of $\gamma_{ij}$, and $\bar{\Gamma}_{kl}^i$ and $\bar{D}$ the Christoffel symbols
and connection of the AdS background 3-metric $\bar{\gamma}_{ij}$. A well-known gauge used in the literature is the harmonic gauge
\cite{DeTurck83,Headrick10,Figueras11,Dias16b}, which is basically a 4D version of \Eref{V}. In \ref{DeTurck}, we clarify the
connection between the two gauges.

To put the first order perturbative geon in the Andersson-Moncrief gauge, we proceed in two steps. First, we infinitesimally
change the time coordinate $t \leftarrow t + \alpha(x^i)$ while keeping unchanged the spatial ones. The transformation rules for
$g_{\alpha\beta}$ at first order in $\alpha$ give
\begin{equation*}
\fl   N \leftarrow N (1 + \beta^i\partial_i \alpha),  \quad
   \beta^i \leftarrow \beta^i + (N^2 \gamma^{ij} + \beta^i\beta^j)\partial_j\alpha , \quad
   \gamma_{ij} \leftarrow \gamma_{ij} - \beta_i \partial_j\alpha - \beta_j \partial_i\alpha .
\end{equation*}
\Eref{K} can be enforced by solving :
\begin{equation}
   2N^2\partial^i\partial_i\alpha + 2\partial_i(N^2\gamma^{ij})\partial_j\alpha + N^2\partial^k\alpha\gamma^{ij}\partial_k\gamma_{ij} - 2NK(1 + \beta^j\partial_j\alpha) = 0,
   \label{eqK}
\end{equation}
where all the geometrical quantities refers to the original coordinates.

Second, we perform an infinitesimal transformation of the spatial coordinates $x^i\leftarrow x^i + \xi^i(x^j)$ while keeping the coordinate $t$ unchanged.
This doesn't change the foliation, so $K$ is left untouched. At first order, the transformation rules for $g_{\alpha\beta}$ give
\begin{eqnarray*}
   N \leftarrow N - \mathcal{L}_{\xi} N , \quad
   \beta^i \leftarrow \beta^i - \mathcal{L}_{\xi}\beta^i , \quad
   \gamma_{ij} \leftarrow \gamma_{ij} - \mathcal{L}_{\xi}\gamma_{ij} , \quad\\
   \Gamma_{kl}^i \leftarrow \Gamma_{kl}^i - \mathcal{L}_\xi \Gamma_{kl}^i - \partial_k \partial_l \xi^i ,
\end{eqnarray*}
where $\mathcal{L}_{\xi}$ stands for the Lie derivative along $\xi$. \Eref{V} is now equivalent to :
\begin{equation}
   - \partial^j\partial_j \xi^i - \gamma^{kl}\mathcal{L}_{\xi}\Gamma_{kl}^i - (\Gamma_{kl}^i - \bar{\Gamma}_{kl}^i)\mathcal{L}_{\xi}\gamma^{kl} + V^i = 0,
   \label{eqV}
\end{equation}
where, again, the geometrical quantities are the ones in the original coordinates.

Equations \eref{eqK} and \eref{eqV} can be solved numerically with the Kadath library. This allows us to compute first order geons in the
Andersson-Moncrief gauge.

\subsection{Solving Einstein equations in the Andersson-Moncrief gauge}

We recall the 3+1 equations with cosmological constant

\numparts
\begin{eqnarray}
   \label{ham}
   R + K^2 - K_{ij}K^{ij} - 2\Lambda = 0 ,\\
   \label{mom}
   D_j K^j_i - D_i K = 0 ,\\
   \label{evo}
   -\mathcal{L}_{m} K_{ij} - D_iD_jN + N(R_{ij} + KK_{ij} - 2K_{ik}K^k_j - \Lambda \gamma_{ij}) = 0 ,
\end{eqnarray}
\endnumparts
where $R_{ij}$ is the Ricci tensor of the 3-metric $\gamma_{ij}$.

First, in \cite{Andersson03}, it was shown that
\begin{equation}
   R_{ij} = -\frac{\gamma^{kl}}{2}\bar{D}_k \bar{D}_l\gamma_{ij} + D_{(i}V_{j)} + \tn{terms in }\partial\gamma\partial\gamma,
\end{equation}
i.e. that, as far as second derivatives are concerned, the Ricci tensor can be decomposed into a well-posed Laplacian-like
operator plus a term $D_{(i}V_{j)}$. Let us mention that this is similar to the properties of the Dirac gauge when a conformal
decomposition of the spatial metric is performed \cite{Bonazzola04}. We then write a 3+1 Einstein-Andersson-Moncrief system
by replacing in \eref{ham}-\eref{evo} all occurrences of $K$ by zero, as is customary for maximal slicing, and all occurrences of
$R_{ij}$ by $R_{ij} - D_{(i}V_{j)}$, leading to
\numparts
\begin{eqnarray}
   \label{ham2}
   R - D_i V^i - K_{ij}K^{ij} - 2\Lambda = 0,\\
   \label{mom2}
   D_j K^j_i = 0, \\
   \label{evo2}
   -\mathcal{L}_{m} K_{ij} - D_iD_jN + N(R_{ij} - D_{(i}V_{j)} - 2K_{ik}K^k_j - \Lambda \gamma_{ij}) = 0 .
\end{eqnarray}
\endnumparts

Unlike the original system (equations \eref{ham}-\eref{evo}), this one is invertible. However one needs to check, a posteriori,
that the obtained solution satisfy the gauge conditions $K=0$ and $V^i=0$. In \cite{Dias16b} the same kind of technique is used to
enforce four-dimensional harmonic gauge, leading to the so-called De Turck method (see \Sref{DeTurck} for a comparison of the two
gauges).

\section{Regularization}
\label{reg}

We chose to work in the so-called conformal coordinates, in which the AdS length element takes the form
\begin{equation}
\fl   \bar{ds}^2 = \bar{g}_{\alpha\beta}dx^\alpha dx^\beta =  -\left(\frac{1+\rho^2}{1 - \rho^2}\right)^2dt^2 +
\frac{4}{(1-\rho^2)^2}(dx^2+dy^2+dz^2), \quad \rho = r/L,
\label{confcoor}
\end{equation}
where $r = \sqrt{x^2 + y^2 + z^2} \in [0,L]$.

Defining the conformal factor
\begin{equation}
   \Omega = \frac{1-\rho^2}{1+\rho^2},
\end{equation}
it is clear that the AdS metric diverges at the boundary $r=L$ like $O(\Omega^{-2})$. We then introduce the conformal background
metric
\begin{equation}
   \hat{\bar{ds}}^2 \equiv \Omega^2 \bar{ds}^2 = \hat{\bar{g}}_{\alpha\beta}dx^\alpha dx^\beta =  -dt^2 + \frac{4}{(1+\rho^2)^2}(dx^2+dy^2+dz^2),
\end{equation}
which is regular and flat at $r = L$. This makes the conformal metric much better suited for numerical computations than the
physical diverging one. Hereafter, we denote by a hat all geometrical quantities that we regularize using some power of $\Omega$,
such that all hatted quantities are regular. In \Tref{quantities} the behavior of various geometric quantities and their
conformal regularization are summarized.

\begin{table}[h]
\centering
\begin{tabular}{lc|lc}
\hline
\hline
Quantity & behavior at $r=L$ & Regularization & behavior at $r=L$\\
\hline
\hline
$g_{\alpha\beta}$ & $O(\Omega^{-2})$ & $\hat{g}_{\alpha\beta} \equiv \Omega^2 g_{\alpha\beta}$ & $O(1)$  \\
$N$               & $O(\Omega^{-1})$ & $\hat{N} \equiv \Omega N$                               & $O(1)$  \\
$\beta^i$         & $O(1)$           & $\hat{\beta}^i \equiv \beta^i$                          & $O(1)$  \\
$\gamma_{ij}$     & $O(\Omega^{-2})$ & $\hat{\gamma}_{ij} \equiv \Omega^2 \gamma_{ij}$         & $O(1)$  \\
$\gamma^{ij}$     & $O(\Omega^{2})$  & $\hat{\gamma}^{ij} \equiv \gamma^{ij}/\Omega^2$         & $O(1)$  \\
$\Gamma_{ij}^k$   & $O(\Omega^{-1})$ & $\hat{\Gamma}_{ij}^k \equiv \Omega\Gamma_{ij}^k$        & $O(1)$  \\
$K_{ij}$          & $O(\Omega^{-2})$ & $\hat{K}_{ij} \equiv \Omega^2K_{ij}$                    & $O(1)$  \\
$K^j_i$           & $O(1)$           & $\hat{K}^j_i \equiv K^j_i$                              & $O(1)$  \\
$K$               & $O(1)$           & $\hat{K} \equiv K$                                      & $O(1)$  \\
$K^{ij}$          & $O(\Omega^{2})$  & $\hat{K}^{ij} \equiv K^{ij}/\Omega^2$                   & $O(1)$  \\
$R_{ij}$          & $O(\Omega^{-2})$ & $\hat{R}_{ij} \equiv \Omega^2 R_{ij}$                   & $O(1)$  \\
$R$               & $O(1)$           & $\hat{R} \equiv R$                                      & $O(1)$  \\
$V^i$             & $O(\Omega)$      & $\hat{V}^i \equiv V^i/\Omega$                           & $O(1)$  \\
$V_i$             & $O(\Omega^{-1})$ & $\hat{V}_i \equiv \Omega V_i$                           & $O(1)$  \\
\hline
\end{tabular}
\caption{Behavior of geometrical quantities near the AdS boundary expressed in powers of the conformal factor $\Omega$.}
\label{quantities}
\end{table}

It is then straightforward to show that the left-hand sides of equations \eref{ham2}-\eref{evo2} behave respectively like $O(1)$,
$O(\Omega^{-1})$ and $O(\Omega^{-3})$ at the AdS boundary. We then denote $\Omega_i = \partial_i \Omega$ and introduce a
regularized $O(1)$ version of the system:
\numparts
\begin{eqnarray}
   \label{hamreg}
   \hat{R} - \partial_i (\Omega \hat{V}^i) - \hat{\Gamma}_{ij}^i \hat{V}^j - \hat{K}_{ij}\hat{K}^{ij} - 2\Lambda = 0,\\
   \label{momreg}
   \Omega \partial_j \hat{K}^j_i + \hat{\Gamma}_{jk}^j \hat{K}^k_i - \hat{\Gamma}_{ij}^k \hat{K}^j_k = 0, \\
   \label{evoreg}
   -\Omega\mathcal{L}_{m} \hat{K}_{ij} - 2 \hat{K}_{ij}\hat{\beta}^k \Omega_k - \delta \hat{N}_{ij}\\
\nonumber+ \hat{N}(\hat{R}_{ij} - \Omega \partial_{(i}\hat{V}_{j)} + \hat{V}_{(i}\Omega_{j)} + \hat{\Gamma}_{ij}^k \hat{V}_k - 2\hat{K}_{ik}\hat{K}^k_j - \Lambda \hat{\gamma}_{ij}) = 0,
\end{eqnarray}
\endnumparts
where the following regularizations hold
\numparts
\begin{eqnarray}
\fl   \delta \hat{N}_{ij} &\equiv& \Omega^3 D_i D_j \left( \frac{\hat{N}}{\Omega} \right) \\
\fl\nonumber   &=& \Omega^2 \partial_i\partial_j \hat{N} - \Omega(\Omega_i \partial_j \hat{N} + \Omega_j \partial_i \hat{N} + \hat{N} \partial_i \partial_j \Omega + \hat{\Gamma}_{ij}^k \partial_k \hat{N}) + 2\hat{N}\Omega_i\Omega_j + \hat{N}\hat{\Gamma}_{ij}^k \Omega_k,\\
\fl   \hat{K}_{ij} &\equiv& \Omega^2 K_{ij} = -\frac{1}{2 \hat{N}}(\Omega \mathcal{L}_m \hat{\gamma}_{ij} + 2 \hat{\gamma}_{ij}\hat{\beta}^k\Omega_k),\\
\fl   \hat{\Gamma}_{ij}^k &\equiv& \Omega \Gamma_{ij}^k = \frac{\Omega}{2}\hat{\gamma}^{kl}(\partial_i \hat{\gamma}_{jl} + \partial_j \hat{\gamma}_{il} - \partial_l \hat{\gamma}_{ij}) - \hat{\gamma}^{kl}(\hat{\gamma}_{il}\Omega_j + \hat{\gamma}_{jl}\Omega_j - \hat{\gamma}_{ij}\Omega_l),\\
\fl   \hat{R}_{ij} &\equiv& \Omega^2 R_{ij} = \Omega (\partial_k \hat{\Gamma}_{ij}^k - \partial_i \hat{\Gamma}_{jk}^k) - (\hat{\Gamma}_{ij}^k\Omega_k - \hat{\Gamma}_{jk}^k\Omega_i) + \hat{\Gamma}_{ij}^k \hat{\Gamma}_{kl}^l - \hat{\Gamma}_{jk}^l \hat{\Gamma}_{il}^k .
\end{eqnarray}
\endnumparts

System \eref{hamreg}-\eref{evoreg} is then a regularized 3+1 Einstein-Andersson-Moncrief system in asymptotically AdS spacetimes. This is the system of
relevance for our numerical computations.

Incidentally, equations \eref{eqK} and \eref{eqV} have to be regularized too for numerical needs (see \Sref{gaugereg}).

\section{Asymptotically AdS spacetimes}
\label{aads}

To give a precise definition of AdS asymptotics, we refer\footnote{Beware that in our notations, $g_{\alpha\beta}$ denotes the
physical metric and $\hat{g}_{\alpha\beta}$ denotes the conformal one while in \cite{Ashtekar84,Ashtekar00} the opposite
convention is chosen.} to \cite{Ashtekar00,Ashtekar84}. A necessary conditions for a spacetime to be asymptotically AdS is to
have a Weyl tensor that vanishes at the boundary. As the Weyl tensor is a conformal invariant, it can be computed either with the
physical metric $g_{\alpha\beta}$ or with its conformal counterpart $\hat{g}_{\alpha\beta}$:
\begin{equation}
   C^\alpha_{\beta\mu\nu} = \hat{\mathcal{C}}^\alpha_{\beta\mu\nu} \equalhat 0,
\end{equation}
where $C^\alpha_{\beta\mu\nu}$ is the Weyl tensor of $g_{\alpha\beta}$, $\hat{\mathcal{C}}^\alpha_{\beta\mu\nu}$ that of
$\hat{g}_{\alpha\beta}$ and $\equalhat$ means equality restricted to the AdS boundary $r=L$. However, in four dimensions, the
vanishing of the Weyl tensor is not a strong enough condition to ensure the AdS asymtotics. A sufficient condition is then to
require its leading order magnetic part to vanish, namely
\begin{equation}
   \hat{B}_{\alpha\beta} \equiv \frac{1}{\Omega}*\hat{\mathcal{C}}_{\alpha\mu\beta\nu}\hat{\nabla}^\mu \Omega \hat{\nabla}^\nu\Omega \equalhat 0,
\end{equation}
where $\hat{\nabla}$ is the connection of $\hat{g}_{\alpha\beta}$ and $*$ denotes Hodge duality.

This definition comes with conserved charge definitions. Given an asymptotically conformal Killing vector $\hat{\xi}^\alpha$, an
associated conserved charge is obtained with\footnote{AMD = Ashtekar-Magnon-Das}
\begin{equation}
   Q^{AMD}_{\xi}[\Sigma_t] \equiv -\frac{L^3}{8\pi G}\oint_{\partial \Sigma_t} \frac{1}{\Omega} \hat{\mathcal{C}}_{\alpha\mu\beta\nu}\hat{\nabla}^\mu\Omega \hat{\nabla}^\nu\Omega \hat{\xi}^\alpha \hat{u}^\beta \sqrt{\hat{\sigma}} d^2 y,
\end{equation}
where $\hat{\sigma}_{\alpha\beta}$ is the metric induced by $\hat{g}_{\alpha\beta}$ on $\partial \Sigma_t$ (i.e. the 2-sphere
$r=L$) and $\hat{u}^\alpha$ is the unit normal vector to $\Sigma_t$ with respect to $\hat{g}_{\alpha\beta}$. Choosing $\xi^\alpha$
to be either $\partial_t^\alpha$ or $\partial_\varphi^\alpha$ gives
then a numerical measure of the mass $M^{AMD}$ and of the angular momentum $J^{AMD}$ of geons.

There is an other definition of conserved charge detailed in \cite{Balasubramanian99}. It is related to the stress tensor of the
dual CFT. Consider the hypersurfaces $r = \tn{cst}$ and denote by $q_{\alpha\beta}$ the metric induced by $g_{\alpha\beta}$ and by
$\Theta_{\alpha\beta}$ the associated extrinsic curvature. A quasilocal stress tensor can then be defined as
\begin{equation}
   T^{CFT}_{\alpha\beta} = \frac{1}{8\pi G}\left( \Theta_{\alpha\beta} - \Theta q_{\alpha\beta} - \frac{2}{L}q_{\alpha\beta} + L \mathcal{G}_{\alpha\beta} \right),
   \label{tcft}
\end{equation}
where $\mathcal{G}_{\alpha\beta}$ is the Einstein tensor of $q_{\alpha\beta}$. At first sight, one may think that
$T^{CFT}_{\alpha\beta}$ behaves as $O(\Omega^{-2})$ near the AdS boundary, but it is actually a $O(\Omega^2)$ for pure AdS, and a
$O(\Omega)$ for asymptotically AdS solutions (the physical stress tensor of the CFT is actually given by
$T^{CFT}_{\alpha\beta}/\Omega$ at $r = L$). In \Sref{tcftreg}, we explain how it can be regularized and computed numerically.

Given an asymptotically Killing vector $\xi^\alpha$, an associated conserved charge is\footnote{BK = Balasubramanian-Kraus}
\begin{equation}
   Q^{BK}_{\xi}[\Sigma_t] = \oint_{\partial \Sigma_t}T^{CFT}_{\mu\nu}u^\mu \xi^\nu \sqrt{\sigma} d^2 y,
\end{equation}
where $\sigma_{\alpha\beta}$ is the metric induced by $g_{\alpha\beta}$ on $\partial \Sigma_t$ and $u^\alpha$ is the unit normal
vector to $\Sigma_t$ with respect to $g_{\alpha\beta}$. Since $\sqrt{\sigma}= O(\Omega^{-2})$ and $u^\mu = O(\Omega)$ near the AdS
boundary, it is clear that a charge can exist if and only if $T^{CFT}_{\alpha\beta} = O(\Omega)$. Choosing $\xi^\alpha$ equal to
$\partial_t^\alpha$ or $\partial_\varphi^\alpha$ provides us with a second, independent measure of mass $M^{BK}$ and angular
momentum $J^{BK}$ of geons.

Furthermore, it was demonstrated in \cite{Ashtekar00} (see also \cite{Mivskovi09}) that these two definitions of charge are
equivalent in any four-dimensional asymptotically AdS spacetime. In \Sref{kerr}, we test this assumption and our numerics with the
Kerr-AdS metric.

\section{Numerical setup}
\label{numproc}

\subsection{Numerical algorithm}

In the present work, differential equations are solved using the open source KADATH library \cite{Grandclement10}, which provides
a C++ interface for solving relativistic systems of equations with multi-domain spectral methods. This user-friendly library has
been successfully used in a wide range of context (from binary black holes \cite{Uryu12} to boson stars \cite{Grandclement14} to
give a few) certifying its robustness. The library manages non-linear systems with a Newton-Raphson scheme.

In order to construct non-linear numerical geons, we proceed as follows :
\begin{itemize}
   \item We analytically construct a helically symmetric first order geon with the results of \Sref{pert},
      and transform the perturbative solution to be expressed in
      corotating conformal coordinates in which the helical Killing vector is just driven by our time coordinate $t'$
      \begin{equation}
         \partial_{t'}^\alpha = \partial_t^\alpha + \frac{\omega}{m}\partial_{\varphi}^\alpha,
      \end{equation}
      such that $\partial_{t'} g_{\alpha\beta} = 0$ and $\mathcal{L}_m = -\mathcal{L}_{\hat{\beta}}$ in \eref{hamreg}-\eref{evoreg}.
   \item Choosing a suitably small amplitude, the linearized first order geon is further processed numerically to be expressed in the
      Andersson-Moncrief Gauge. This is achieved by solving \Eref{eqKreg} and \Eref{eqVreg} with the Kadath library, as explained
      in \Sref{gauge} and \Sref{gaugereg}.
   \item The resulting first order geon in the Andersson-Moncrief gauge is then used as an initial guess for the full 3+1
      Einstein-Andersson-Moncrief system of ten equations \eref{hamreg}-\eref{evoreg} whose ten unknowns are
      $N$,$\beta^i$,$\gamma_{ij}$. The boundary conditions at the AdS boundary are the following :
      \begin{equation}
         \hat{N} \equalhat \hat{\bar{N}},\quad \hat{\beta}^i \equalhat \frac{\omega}{m}\partial_{\phi}^i, \quad \hat{\gamma}_{ij} \equalhat \hat{\bar{\gamma}}_{ij}.
         \label{BC}
      \end{equation}
      The condition on the shift just translates that the frame is corotating with the geon. As $\omega$ is expected to change
      with the geon amplitude, it is treated as an additional unknown of the system, while we provide an additional equation that
      enforces the marching parameter, or geon wiggliness $w$, to take some user-defined value. The Newton-Raphson algorithm of
      the Kadath library is then in charge of finding the solution. The Newton-Raphson iteration is stopped when the error,
      measured as the highest coefficient of the Einstein equation residuals, reaches about $10^{-8}$.
   \item Once a numerical and non-linear solution is obtained, it is used as an initial guess for the Einstein-Andersson-Moncrief system with a
      $w$ slightly incremented. The system then converges to the nearby solution of the system with this
      wiggliness requirement. Iterating the process, we are able to build sequences of geons parametrized by $w$,
      which represents the amplitude of the non-linear geon.
\end{itemize}

\subsection{Precision monitoring}
\label{prec}
In order to monitor the precision of our numerical results, various tests are performed :

\begin{itemize}
   \item[1-] \textbf{Spectral convergence :} if the metric components are well described by the spectral expansion, their spectral
      coefficients should decrease exponentially. With double precision arithmetics and with a second order differential system of
      equations, the saturation level is expected to be around $10^{-10}$.
   \item[2-] \textbf{Gauge residual :} $K$ and $V^i$ should be as low as possible (but are expected to saturate at a $10^{-10}$
      level). Their infinity norm should decrease with numerical resolution. An other complementary check in the Einstein-Andersson-Moncrief framework
      is to observe a similar convergence for the components of $R_{\alpha\beta} - \Lambda g_{\alpha\beta}$ which should be zero
      for any solution of Einstein equations in vaccum.
   \item[3-] \textbf{Asymptotically AdS spacetimes :} we enforce Dirichlet boundary conditions on the system, however this might
      not be enough to ensure the right asymptotics. According to \Sref{aads}, we can check on one hand that
      $\hat{C}_{\alpha\beta\mu\nu}$, $\hat{B}_{\alpha\beta}$ and $T^{CFT}_{\alpha\beta}$ have boundary values 
      decreasing to zero, and on the other hand that AMD and BK charges converge to each other
      when increasing numerical resolution.
   \item[4-] \textbf{Agreement with perturbative approach :} any numerical sequence of geon should coincide with perturbative
      results for low enough amplitudes.
\end{itemize}

\section{Results}
\label{res}

\subsection{Geons with $(l,m,n) = (2,2,0)$}

Following \Sref{numproc}, we start building geons with excitation number $(l,m,n) = (2,2,0)$, i.e. helical geons with lowest
excitations numbers. We are able to reach unprecedentedly high amplitudes, with deviations from third order perturbative expansion
as large as $50\%$. We will use this family of geons as a testbed for our numerics. Defining
\begin{equation}
   h_{\alpha\beta} \equiv g_{\alpha\beta} - \bar{g}_{\alpha\beta} \quad \tn{and} \quad \hat{h}_{\alpha\beta} \equiv \Omega^2 h_{\alpha\beta},
\end{equation}
a relevant marching parameter, or wiggliness, seems to be
\begin{equation}
   w \equiv \hat{h}_{xx}(r=0) = h_{xx}(r=0),
\end{equation}
in the Andersson-Moncrief gauge, as $h_{xx}$ has a bell shape (see \Fref{fig:gallery}). The sequence typically starts at $w = 0.1$
and finishes at $w = 10$ (compare to the AdS background $\hat{\bar{g}}_{xx}(0) = 4$). We use two domains, one nucleus describing
$r\in[0,0.5]L$ and one shell describing $r\in[0.5,1]L$. We do so in order to compute $\Theta_{\alpha\beta}$ involved in
\Eref{tcft}, because as it diverges like $O(r^{-1})$ near the origin, we can only compute it in the shell domain. As $l$ and $m$
are even, there is an octant symmetry. Accordingly, quoting a resolution of, say ``37x9x9'', means that, in each domain, one uses
$N_r = 37$ points in the $r$ coordinate and $N_\theta = N_\varphi = 9$ points per octant in $\theta$ and $\varphi$ coordinates. We
checked that the results are insensitive to the positioning of the domain separation, as expected for the global representation of
smooth fields in spectral methods.

\begin{figure}[t]
   \centering
      \subfigure[]{\includegraphics[width = 0.49\textwidth]{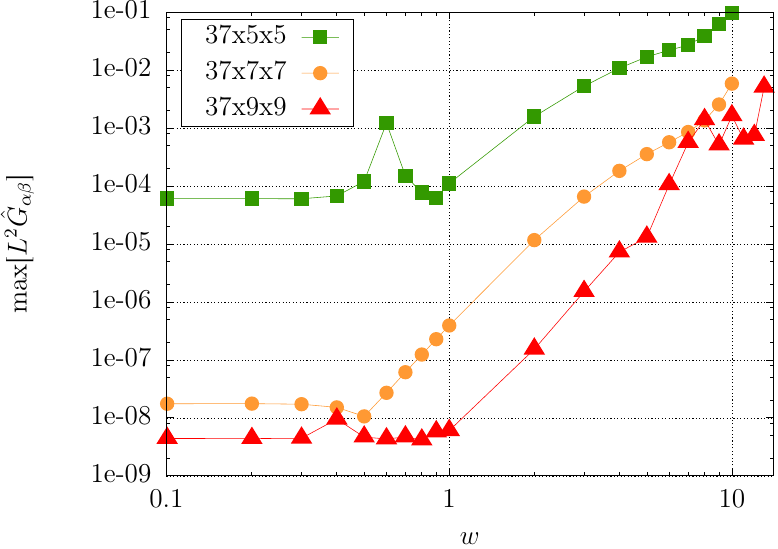}}
      \subfigure[]{\includegraphics[width = 0.49\textwidth]{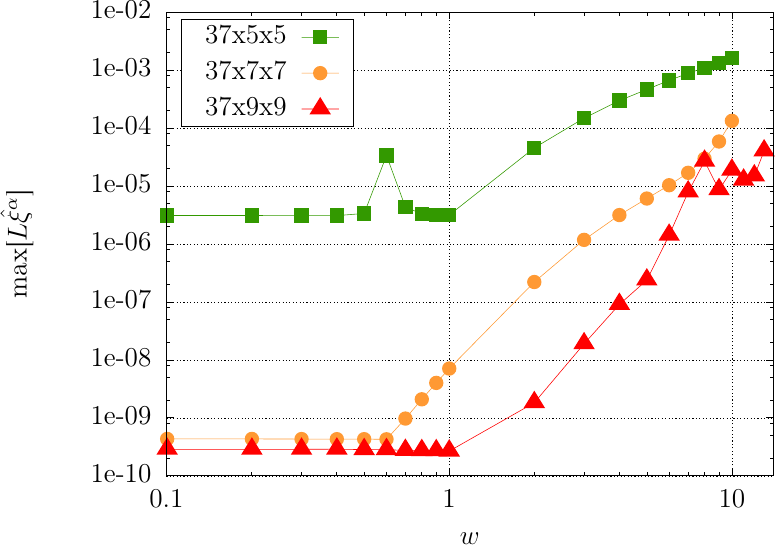}}
      \subfigure[]{\includegraphics[width = 0.49\textwidth]{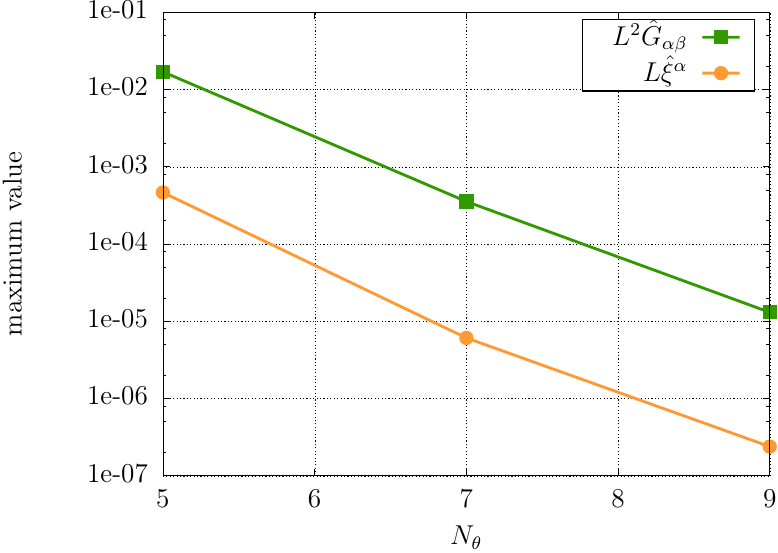}}
      \caption{Top left panel : residual of the regularized Einstein equation tensor
      $\hat{G}_{\alpha\beta} = \hat{R}_{\alpha\beta} - \Lambda \hat{g}_{\alpha\beta}$. Top right panel : residual of the regularized
      gauge vector $\hat{\xi}^\alpha = (-\hat{K},\hat{V}^i)$. Bottom panel: spectral convergence of the Einstein and gauge
      residuals as a function of the angular resolution at fixed wiggliness $w = 5$. Each residual is the maximum value in the
   whole configuration space, i.e. at collocation points, for the $(l,m,n) = (2,2,0)$ geon.}
   \label{fig:eingauge22}
\end{figure}

In \Fref{fig:eingauge22}, we show how the Einstein and gauge residuals of the system of equations vary with amplitude and
resolution. As curves happen to be almost insensitive to radial resolution $N_r \in [29,37]$, we only show the angular resolution
dependence. The errors are increasing with the amplitude of the geon, but decrease exponentially by several orders of magnitude with
resolution, indicating spectral convergence. Thus our solutions are not only solutions of the Einstein-Andersson-Moncrief system
but also of the full Einstein system. At a resolution of 37x9x9, we lower the Einstein residual down to $\sim 10^{-3}$ when the
largest metric coefficient is $\sim 15$ at the origin.

\begin{figure}[t]
   \centering
   \subfigure[]{\includegraphics[width = 0.49\textwidth]{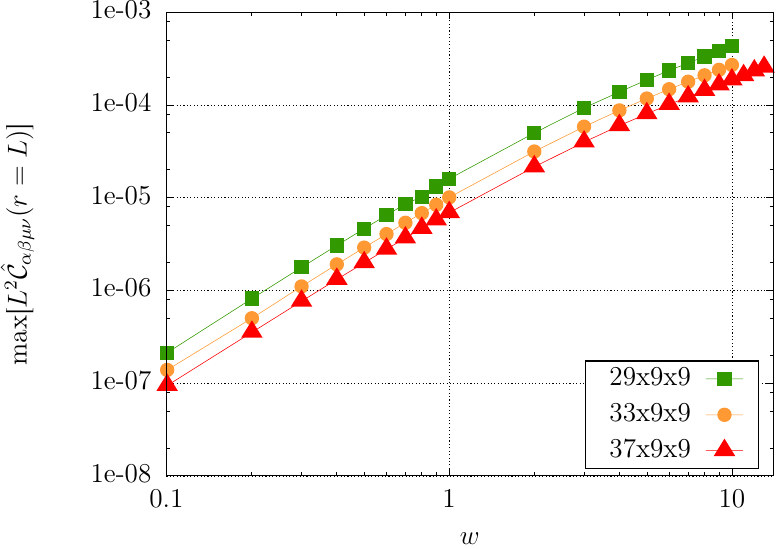}}
   \subfigure[]{\includegraphics[width = 0.49\textwidth]{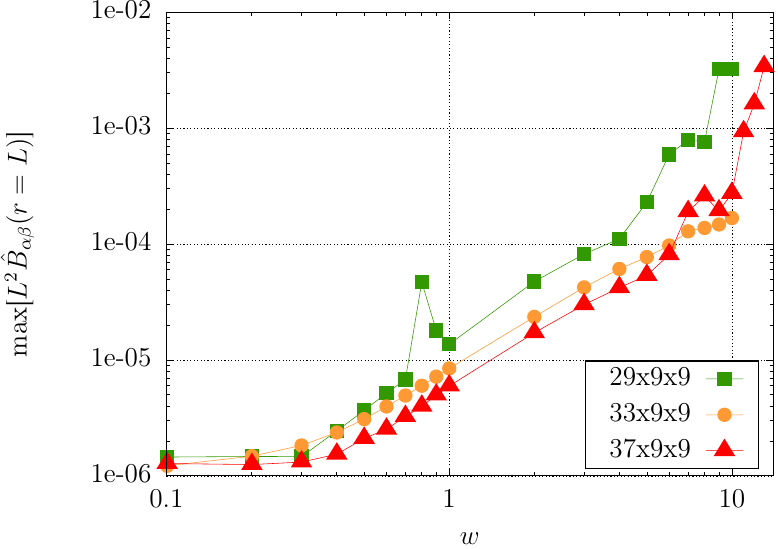}}
   \subfigure[]{\includegraphics[width = 0.49\textwidth]{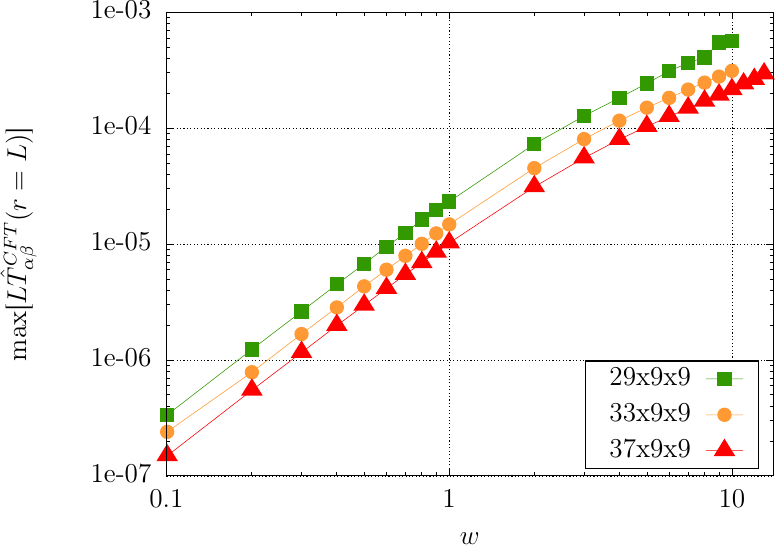}}
   \subfigure[]{\includegraphics[width = 0.49\textwidth]{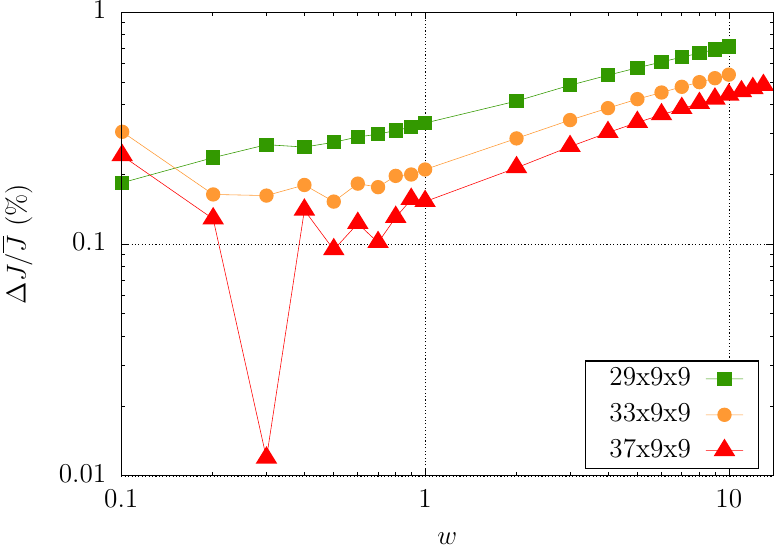}}
   \caption{AdS residuals for the $(l,m,n) = (2,2,0)$ geon. Upper left panel : residual of the boundary Weyl tensor
      $\hat{\mathcal{C}}_{\alpha\beta\mu\nu}$ of $\hat{g}_{\alpha\beta}$. Upper right panel : residual of its rescaled magnetic
      part $\hat{B}_{\alpha\beta}$. Bottom left panel : residual of the CFT stress tensor $T^{CFT}_{\alpha\beta}$. All residuals
      are restricted to the AdS boundary. Bottom right panel : relative difference between AMD and BK angular momenta, with
      $\Delta J = J^{AMD} - J^{BK}$ and $\bar{J} = (J^{AMD} + J^{BK})/2$.}
   \label{fig:aads22}
\end{figure}

In \Fref{fig:aads22}, four AdS asymptotics indicators are evaluated (see \Sref{aads} above). For these, the angular resolution
has essentially no effect for $N_{\theta,\varphi}\in[5,9]$, so we show the radial resolution declination only. The higher the
radial resolution, the closer to zero they are, which shows that our solutions are well asymptotically AdS. The bottom right panel
shows that our AMD and BK charges agree with each other at a $\sim 0.5\%$ level, which is an extra confirmation of the validity of our
solutions. The convergence rate seems to be quite slow, but this is to be expected : these indicators are quite demanding in terms
of precision as they involve second order derivatives, divisions in coefficient space and evaluation or integration at the AdS
boundary.

\begin{figure}[t]
   \centering
   \subfigure[]{\includegraphics[width = 0.49\textwidth]{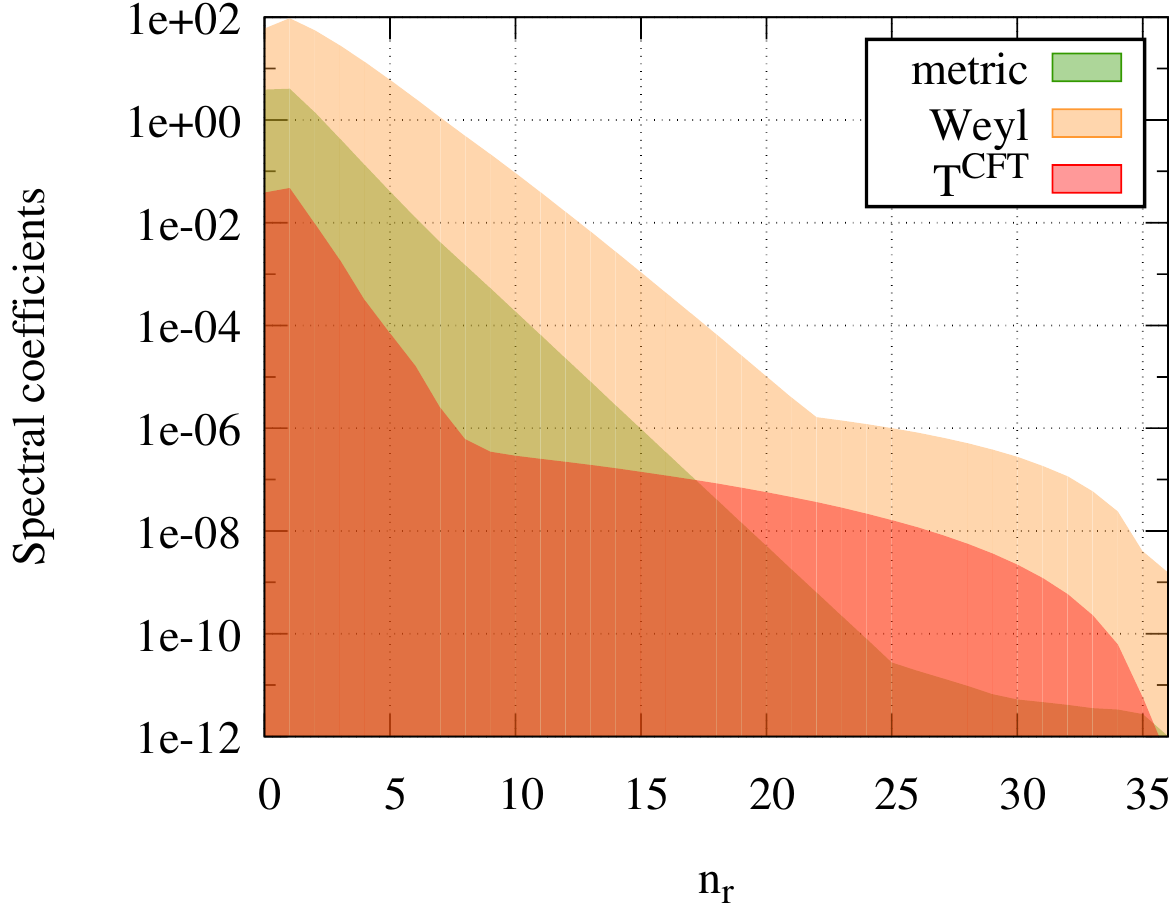}}
   \subfigure[]{\includegraphics[width = 0.49\textwidth]{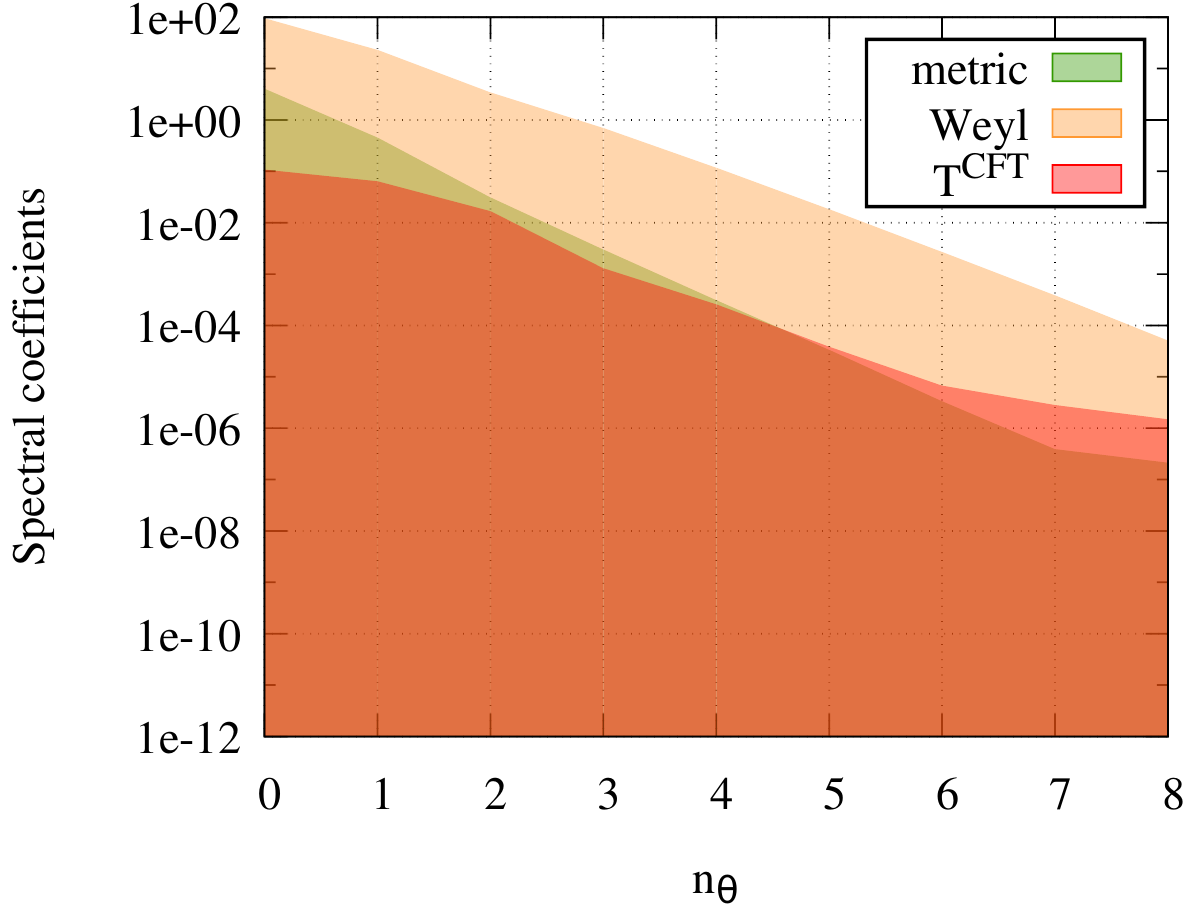}}
   \caption{Spectral coefficients of $\hat{g}_{\alpha\beta}$ in green, $\hat{\mathcal{C}}_{\alpha\beta\mu\nu}$ in yellow and
   $T^{CFT}_{\alpha\beta}$ in red for the $(l,m,n) = (2,2,0)$ geons at a resolution of 37x9x9 and amplitude $w = 10$. Our
   coefficients collection is actually a three-dimensional array indexed by three integers $n_r \in \{0,\cdots,N_r\}$, $n_\theta
   \in \{0,\cdots,N_\theta\}$ and $n_\varphi \in \{0,\cdots,N_\varphi\}$. Left panel : coefficients versus $n_r$ for arbitrary
   $n_\theta$ and $n_\varphi$. Right panel : coefficients versus $n_\theta$ for arbitrary $n_r$ and $n_\varphi$. For the sake of
   clarity, only the upper enveloppe of the coefficients collection is shown.}
   \label{fig:coefs22}
\end{figure}

\Fref{fig:coefs22} shows the coefficients of our solution with largest amplitude $w=10$ and highest resolution 37x9x9. Spectral
convergence is observed both radially and angularly. The saturation threshold is larger for the Weyl tensor
$\hat{\mathcal{C}}_{\alpha\beta\mu\nu}$ and the CFT stress tensor $T^{CFT}_{\alpha\beta}$ as they involve second order derivatives
of the metric and, for the latter, a regularization procedure (\Sref{tcftreg}) that both increase numerical errors and hence noise
level. If it were not too computationally demanding, we could increase the angular resolution to describe better the coefficient
tail, and it would probably decrease the errors on Einstein equation and gauge residuals, as observed in \Fref{fig:eingauge22}.
However, angular resolution doesn't seem critical when it comes to computing global charges.

\begin{figure}[t]
   \centering
   \subfigure[]{\includegraphics[width = 0.49\textwidth]{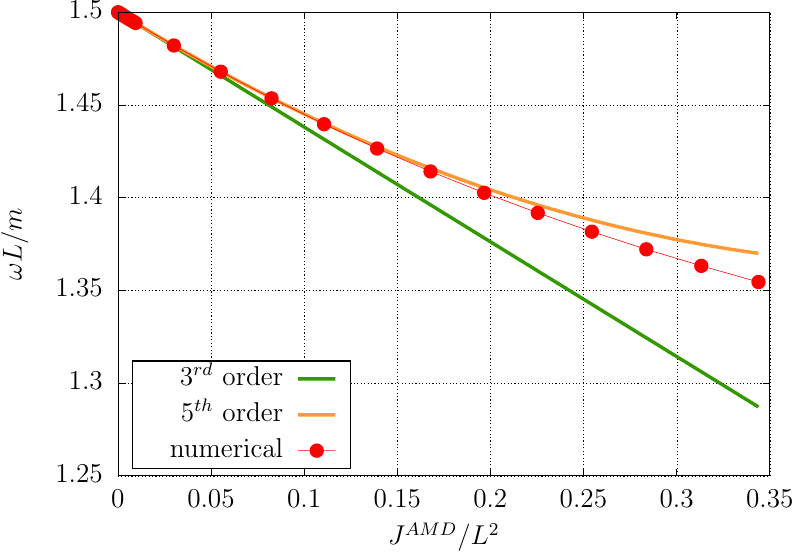}}
   \subfigure[]{\includegraphics[width = 0.49\textwidth]{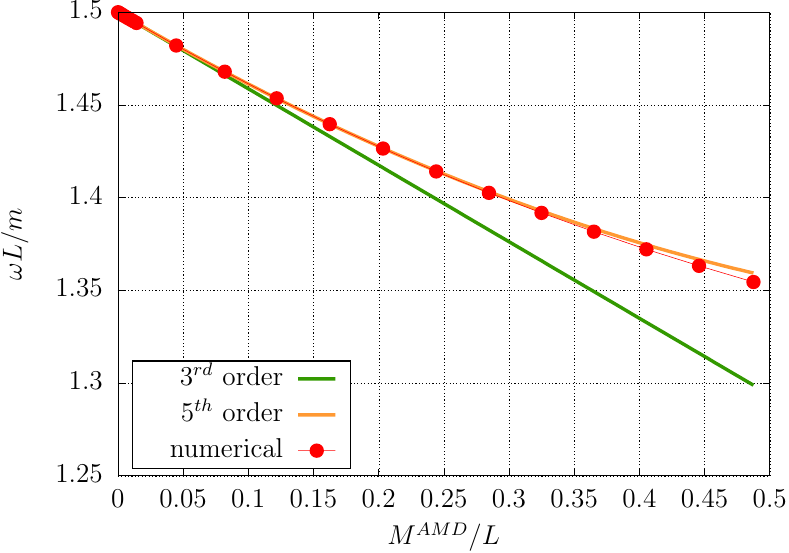}}\\
   \subfigure[]{\includegraphics[width = 0.49\textwidth]{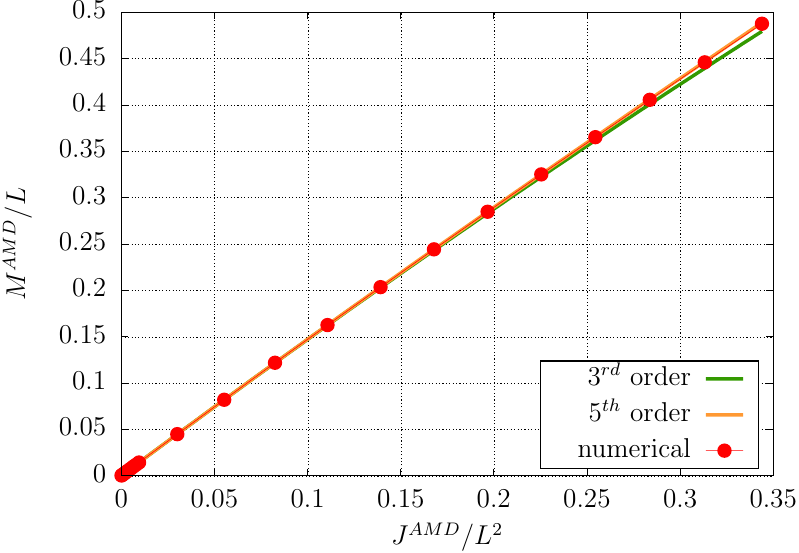}}
   \subfigure[]{\includegraphics[width = 0.49\textwidth]{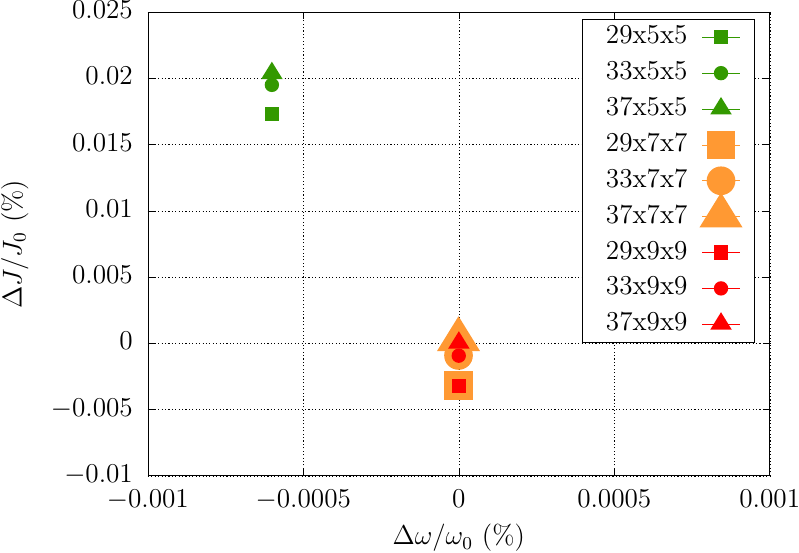}}
   \caption{The $\omega$-$M$-$J$ planes for our numerical sequences $(l,m,n) = (2,2,0)$ at resolution 37x9x9 using AMD
   definitions. The successive perturbative orders are shown in comparison with our highest resolution results. At the bottom
   right panel, we have chosen a typical point on the sequence at resolution 37x9x9 ($w_0 = 5$ , $\omega_0 L/m = 1.44$, $J_0 =
   0.11$). The plot shows the relative difference $\Delta \omega = \omega - \omega_0$ and $\Delta J = J - J_0$ in percentage for
   all resolutions, $w$ being fixed at $w_0$.}
   \label{fig:amd22}
\end{figure}

When it comes to computing geon charges, it turns out that the precision on the mass (be it AMD or BK) is less than that on
the angular momentum. In particular, even if $J$ is in a very strong agreement with perturbative approach in a low amplitude
limit, $M$ is usually overestimated by $\sim5$-$10\%$ depending on the radial resolution. Increasing the number of points improves
the match but very slowly. Our guess is that we lose precision on $M$ because of the numerous steps of regularization and
spectral operations that all bring their own numerical errors which accumulate. This asymmetry between $M$ and $J$ is unclear, but
probably the terms involved in the computation of $J$ are simpler than those involved in the computation of $M$. We think that, if
affordable, quadruple precision could improve this point. So in order to give reliable masses within reasonable computing times,
we compute $M$ using the first law of geon dynamics $\delta M - \omega \delta J/m = 0$, which ensures that $M$ is computed with as
much precision as $\omega$ and $J$ are. This relation is demonstrated in the asymptotically flat case for helically symmetric
system \cite{Friedman02} with respect to the Arnowitt-Deser-Misner global charges, and a similar results holds for Kerr-AdS
\cite{Gibbons05} with an additional entropy term. Let us also mention that a sketch of a proof is present in \cite{Horowitz14}.
But as far as we know, a rigorous mathematical proof in the general helically symmetric case in asymptotically AdS spacetimes is
still missing. Nevertheless the first law is widely accepted and actually confirmed by perturbative results up to sixth order (see
\Eref{mjexpansion}).

In practice, we write
\begin{equation}
   M = \int_0^J \frac{\omega(J')}{m}dJ',
\end{equation}
where the function $\omega(J)$ is obtained by a polynomial fit (reduced $\chi^2 < 10^{-13}$).

\Fref{fig:amd22} displays the three global quantities of importance : the angular velocity $\omega/m$ (with $m=2$), the mass $M$
and angular momentum $J$. It is clear that successive orders of perturbations are closer and closer to our numerical solutions. In
order to estimate the numericall error bars, we examine one single point of the sequence, say $w = 5$, at all
our available resolutions. Taking as reference values the one computed at our highest resolution 37x9x9, we look at the difference
with the lower resolutions results. We naturally expect small differences in the numerical measurements of $\omega$ and $J$
depending on the resolution. These differences are pictured on the bottom right panel of figure \ref{fig:amd22}. First, it is clear that
the results are converging to the highest resolution results (origin of the plot). Second, this allows us to read off error bars on
$J$ and $\omega$, namely $\Delta J \sim 0.02 \%$ and $\Delta \omega = 0.0006 \%$ between our worst and best resolutions.
Restricting the angular resolutions between 7 and 9 angular points (hardly distinguishable on the figure) gives $\Delta J = 0.003 \%$
and $\Delta \omega = 0.000001 \%$. Furthermore, we observed that these error bars remained approximately constants along the entire sequence.
This magnitude of error bars is obviously indistinguishable at naked eye on the three other panels of figure \ref{fig:amd22}.

\begin{figure}[t]
   \centering
   \includegraphics[width = \textwidth]{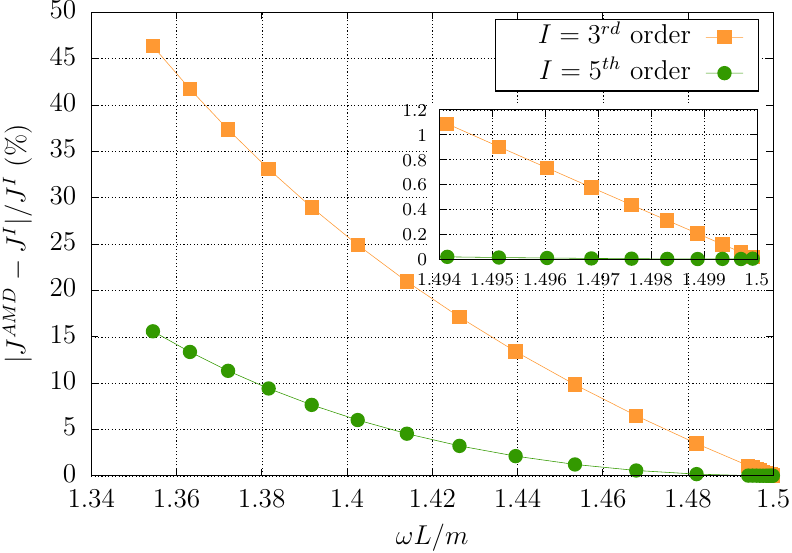}
   \caption{Difference between the numerical $J^{AMD}$ for $(l,m,n) = (2,2,0)$ at resolution 37x9x9 and perturbative approach at
   $I^{th}$ order, $I$ being 3 (yellow squares) or 5 (green circles). Inset : zoom on the low amplitude limit $\omega L/m
\rightarrow 3/2$.}
   \label{fig:4order22}
\end{figure}

In \Fref{fig:4order22}, we show the difference between the numerical results and perturbative predictions. This demonstrates that
the non-linear solutions deviates from $3^{rd}$ order by at most $50\%$ and from $5^{th}$ order by at most $15\%$.

If we consider the expansion $\omega L/m = f(J)$
\begin{equation}
   \frac{\omega L}{m} = a_0 + a_1 \frac{J}{L^2} + a_2 \frac{J^2}{L^4} + a_3 \frac{J^3}{L^6} + \cdots,
   \label{omjexpansion2}
\end{equation}
the $a_i$ coefficients can be computed by a polynomial fit. \Tref{coefs22} shows the coefficients of the perturbative
and numerical results (see \Eref{omjexpansion}).

\begin{table}[h]
\centering
\begin{tabular}{lcccc}
             & $a_0$           & $a_1$           & $a_2$           & $a_3$           \\
\hline
perturbative & $1.5000000$     & $-0.619062$     & $0.70049$       &  -              \\
numerical    & $1.5000000$     & $-0.619064$     & $0.70031$       & $-0.345$        \\
error        & $\pm 6.10^{-9}$ & $\pm 2.10^{-6}$ & $\pm 7.10^{-5}$ & $\pm 1.10^{-3}$ \\
\hline
\end{tabular}
\caption{Coefficients in the polynomial expansion $\omega L/m = f(J)$ for both pertubative and numerical results at resolution
37x9x9. Error bars are given by the Levenberg-Marquardt fit algorithm.}
\label{coefs22}
\end{table}

The numerical and available perturbative values of the coefficients agree very well. For instance the relative difference in $a_2$
is of order $0.01 \%$.

At this point, let us mention that our results are in disagreement with those of \cite{Horowitz14}, whose authors were the first (and
single) to propose a numerical construction of the $(l,m,n) = (2,2,0)$ geons. Indeed, looking at their figure (1.b), it is clear
that they found $a_2$ to be negative. According to our results, listed in table \ref{coefs22}, we find, however, that $a_2\sim
+0.700$. We are very confident in this result as our perturbative computations carried out to the sixth order agree very well with
our precise numerical measurements. We thus provide two independent arguments in favour of the positivity of $a_2$, and are unable
to find any reason why this coefficient should be negative in \cite{Horowitz14}. Additional and independent future derivations of
these results would provide a very welcome clarification of this point.

\subsection{Geons with $(l,m,n) = (4,4,0)$}

\begin{figure}[t]
   \centering
   \includegraphics[width = \textwidth]{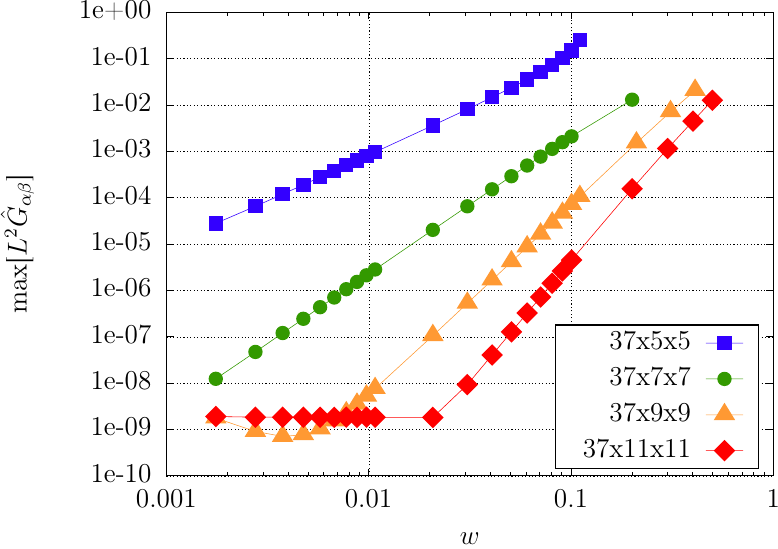}
   \caption{Residual of the regularized Einstein equation tensor $\hat{G}_{\alpha\beta} = \hat{R}_{\alpha\beta} - \Lambda
   \hat{g}_{\alpha\beta}$ for $(l,m,n) = (4,4,0)$.}
   \label{fig:err44}
\end{figure}

Increasing the angular number of excitations, we can construct geons with $(l,m,n) = (4,4,0)$. As a wiggliness parameter, we
choose the largest positive coefficient in the spectral expansion of the first order geon, namely :

\begin{equation}
   w \equiv \tn{coefficient } (n_r = 0,n_\theta = 1,n_\varphi = 1) \tn{ of } h_{yy},
\end{equation}
\Fref{fig:err44} shows the Einstein residuals at different resolutions. The exponential decrease in the errors when increasing the
number of collocation points demonstrates that our solutions are indeed solutions of Einstein equation. Similar plots hold for the
other indicators detailed in the previous subsection.

\begin{figure}[h]
   \centering
   \subfigure[]{\includegraphics[width = 0.49\textwidth]{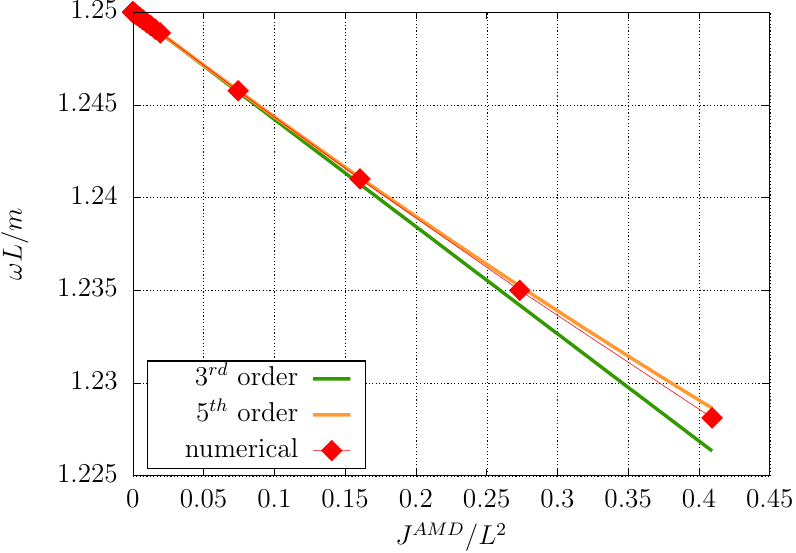}}
   \subfigure[]{\includegraphics[width = 0.49\textwidth]{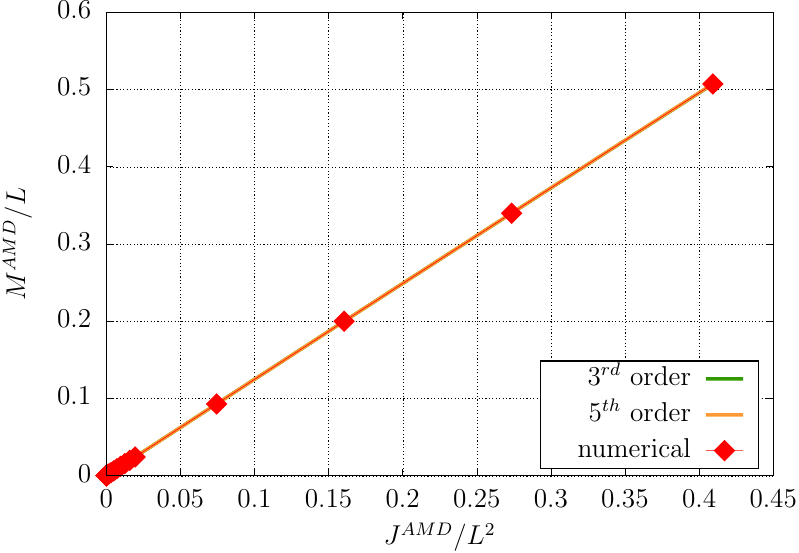}}
   \caption{The $\omega$-$M$-$J$ planes for our numerical sequences $(l,m,n) = (4,4,0)$ at resolution 37x11x11 using AMD
   definitions. On the right panel, the curves are indistinguishable at naked eye.}
   \label{fig:amd44}
\end{figure}

Global quantites are shown on \Fref{fig:amd44}. Our numerical results match the second order perturbative results in the low
amplitude limit. We can reach masses of order $\sim 0.5$ in AdS length units.

Fitting our numerical data, we infer the numerical values of the coefficients in the $\omega L/M = f(J)$ expansion
\eref{omjexpansion2}. They are presented in \Tref{coefs44}. In order to get the coefficients $a_1$ and $a_2$ we had to go to fifth
order in the $\varepsilon$ expansion, obtaining
\numparts
\begin{eqnarray}
\fl \omega_1&=&-\frac{890231}{4900896\pi}\approx-0.0578199,\\
\fl \omega_2&=&\frac{29082107232588401} {1122863180414976} - \frac{777842303655609302903087}{3044556205497398592000\pi^2}\approx 0.0137851 .
\end{eqnarray}
\endnumparts

\begin{table}[h]
\centering
\begin{tabular}{lcccc}
             & $a_0$            & $a_1$            & $a_2$           & $a_3$           \\
\hline
perturbative & $1.2500000$      & $-0.05781990$    & $0.0137851$     &  -              \\
numerical    & $1.2500000$      & $-0.05782040$    & $0.0111142$     & $-0.0011196$    \\
error        & $\pm 4.10^{-10}$ & $\pm 3.10^{-8}$  & $\pm 2.10^{-7}$ & $\pm 4.10^{-7}$ \\
\hline
\end{tabular}
\caption{Coefficients in the polynomial expansion $\omega L/m = f(J)$ for both pertubative and numerical results of $(l,m,n) =
   (4,4,0)$ at resolution 37x11x11. Error bars are given by the Levenberg-Marquardt fit algorithm.}
\label{coefs44}
\end{table}

Unfortunately, it is hard and time-consuming to push the sequence to higher amplitudes, so we cannot predict the successive
coefficients with reasonable precision for the time being.

\subsection{Geons with one radial node}

The perturbative approach for radially excited geons is more complicated. At first, in \cite{Dias16a}, the authors claimed that
the $(l,m,n) = (2,2,1)$ linear mode cannot seed a stable non-linear family of geons, because some secular resonances remain
even after the Poincar\'e-Linstedt regularization. However, recently a paper and a comment came out suggesting that a linear combination of
several seeds sharing the same $\omega$ could indeed survive at arbitrary order \cite{Rostworowski16,Rostworowski17}.

The angular frequency belonging to the helical extension of the $(2,2,1)$ scalar mode is $\bar\omega/m=5/2$, which is the same as
the angular frequency of the scalar mode $(4,2,0)$ and the vector mode $(3,2,0)$. Taking a linear combination of the helically
symmetric metric perturbation generated by these three modes, we obtain a three-parameter seed for the perturbation formalism.
Evaluating the generating function \eref{eqgenfscal} with $l=2$, $n=1$ and $\delta=0$, calculating the gauge invariant variables
using \eref{zscalareqs1}-\eref{zscalareqs4} with zero source terms, then using \eref{eqshhzz} and \eref{scalmetrpert} with $m=2$
we get the class $(2,2,1)$ metric perturbation. Adding the metric perturbation $(2,-2,1)$ with the same amplitude $\alpha$, but
with phase $\delta=\pi/2$, we get the helically symmetric rotating version of it. We repeat the same procedure for the scalar mode
$(4,2,0)$, but with amplitude $\beta$ instead of $\alpha$. The rotating $(3,2,0)$ vector mode, with amplitude $\gamma$, is
obtained similarly, using \eref{eqgenfvect}, \eref{zvectoreqs}, \eref{eqshhzz} and \eref{vectmetrpert}. Using this three parameter
seed metric for $h^{(1)}_{ab}$, at second order in the $\varepsilon$ expansion an unspecified constant $\nu_2$ arises at the
$(l,m)=(0,0)$ mode, according to \eref{nukdefeq}, which describes the change of the oscillation frequency. The second order
perturbation equations always have periodic regular asymptotically AdS solutions, but at third order in $\varepsilon$ three
consistency conditions arise, at the scalar $(l,m)=(2,2)$ and $(4,2)$ modes, and at the $(3,2)$ vector mode. Each of these three
conditions is a long polynomial, they have one term linear in $\nu_2$, and rest of the terms are cubic and homogeneous in
$\alpha$, $\beta$ and $\gamma$. They can be transformed to a $13$-th degree polynomial equation in one variable, which can be
solved only numerically, and has $3$ real and $10$ complex roots. The leading order angular momentum and mass is
\begin{equation}
 J=\frac{5(48\alpha^2+19845\beta^2+245\gamma^2)}{28672L^3}\varepsilon^2, \quad
 M=\frac{5}{2L}J \ . \label{nodesoljm}
\end{equation}
The physical frequency of the solution is $\omega=\bar\omega/\sqrt{\nu}$, where $\nu=L^2(1+\nu_2\varepsilon^2)$, according to
\eref{eqnudef}. It follows that the angular frequency to second order in $\varepsilon$ is
\begin{equation}
 \frac{\omega L}{m}=\frac{5}{2}-\frac{5\nu_2}{4L}\varepsilon^2
 =a_0+a_1\frac{J}{L^2} \ ,
\end{equation}
where from \eref{nodesoljm} we get that
\begin{equation}
 a_0=\frac{5}{2}, \quad
 a_1=-\frac{3584\nu_2 L^4}{48\alpha^2+19845\beta^2+245\gamma^2} \ .
\end{equation}
The numerical values for of ratio of the amplitudes and of the frequency change parameters are given in Table
\ref{threefamtable}.~for the three one-parameter families of solutions that satisfy the consistency conditions at $\varepsilon^3$
order.
\begin{table}[h]
\centering
\begin{tabular}{lcccc}
  & $\displaystyle\frac{\beta}{\alpha}$ & $\displaystyle\frac{\gamma}{\alpha}$ & $\displaystyle\frac{\nu_2L^4}{\alpha^2}$ & $a_1$ \\[2.5mm]
\hline
family I.   & $-0.00286074$ & $0.154618$ & $0.0461014$ & $-3.05866$  \\
family II.  & $0.0367439$   & $-1.67172$ & $0.158401$  & $-0.747498$ \\
family III. & $1.07086$     & $1.39907$  & $1.17356$   & $-0.180638$ \\
\hline
\end{tabular}
\caption{Numerical values of the parameters for the three one-parameter families of solutions with angular frequency $\bar\omega/m=5/2.$}
\label{threefamtable}
\end{table}

In order to get radially excited geons, we start with the combination of the three linear modes with amplitudes ratios given in Table \ref{threefamtable}.
Our marching parameters are
\numparts
\begin{eqnarray}
   w &\equiv& \tn{coefficient } (n_r = 1,n_\theta = 0,n_\varphi = 0) \tn{ of } \hat{h}_{yy} \tn{ for family I},\\
   w &\equiv& \tn{coefficient } (n_r = 0,n_\theta = 1,n_\varphi = 0) \tn{ of } \hat{h}_{xx} \tn{ for family II},\\
   w &\equiv& \tn{coefficient } (n_r = 0,n_\theta = 1,n_\varphi = 0) \tn{ of } \hat{h}_{xz} \tn{ for family III}.
\end{eqnarray}
\endnumparts
We also tried to naively start with a $(l,m,n) = (2,2,1)$ first order seed, and observed that the code was converging to the
family I branch of solutions, as it is the one with highest contribution from this seed. We built numerically all three families
of excited geons.

\begin{figure}[t]
   \centering
   \subfigure[]{\includegraphics[width = 0.49\textwidth]{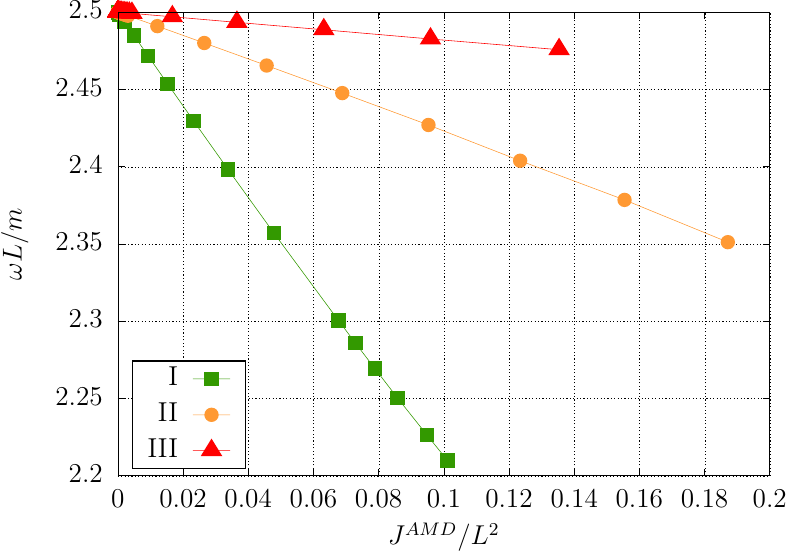}}
   \subfigure[]{\includegraphics[width = 0.49\textwidth]{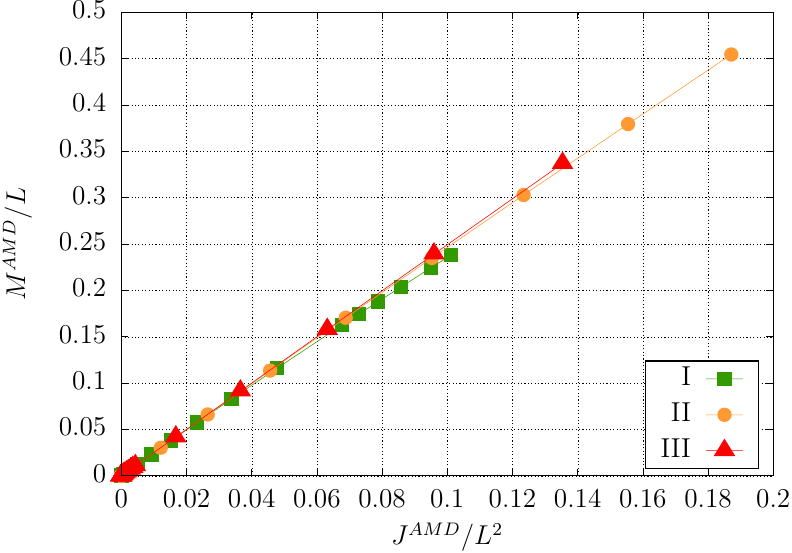}}
   \caption{The $\omega$-$M$-$J$ planes for the three families I,II and III of radially excited geons with one radial node and $m
      = 2$ using AMD definitions. All curves are computed at a resolution of 37x9x9.}
   \label{fig:amd1node}
\end{figure}

Global quantities are displayed on \Fref{fig:amd1node}. The numerical results are in good agreement with
second order perturbative ones in the low amplitude limit. We can reach masses of order $\sim 0.5$ in AdS length unit depending on
the family.

\begin{table}[h]
\centering
\begin{tabular}{llccccc}
                          &              & $a_0$           & $a_1$           & $a_2$           & $a_3$           \\
\hline
\multirow{3}*{family I}   & perturbative & $2.5000000$     & $-3.058658$     &     -           &    -            \\
                          & numerical    & $2.5000000$     & $-3.058670$     & $1.239$         & $8.2$           \\
                          & error        & $\pm 2.10^{-9}$ & $\pm 7.10^{-6}$ & $\pm 2.10^{-3}$ & $\pm 3.10^{-1}$ \\
\hline
\multirow{3}*{family II}  & perturbative & $2.5000000$     & $-0.747498$     &     -           &    -            \\
                          & numerical    & $2.5000000$     & $-0.747533$     & $-0.1723$       & $-0.39$         \\
                          & error        & $\pm 3.10^{-9}$ & $\pm 3.10^{-6}$ & $\pm 3.10^{-4}$ & $\pm 2.10^{-2}$ \\
\hline
\multirow{3}*{family III} & perturbative & $2.5000000$     & $-0.180638$     &     -           &    -            \\
                          & numerical    & $2.5000000$     & $-0.180628$     & $-0.0072$       &    -            \\
                          & error        & $\pm 5.10^{-9}$ & $\pm 5.10^{-6}$ & $\pm 7.10^{-4}$ &    -            \\
\hline
\end{tabular}
\caption{Coefficients in the polynomial expansion $\omega L/m = f(J)$ for both pertubative and numerical results at a resolution
of 37x9x9. Error bars are given by the Levenberg-Marquardt fit algorithm.}
\label{coefs1nodeFamily1}
\end{table}

\Tref{coefs1nodeFamily1} shows the prediction we can make on the expansion coefficients with our numerical results.
We expect that higher amplitudes sequences at higher resolutions could allow us to refine these predictions.

\begin{figure}[t]
   \centering
   \includegraphics[width = 0.98\textwidth]{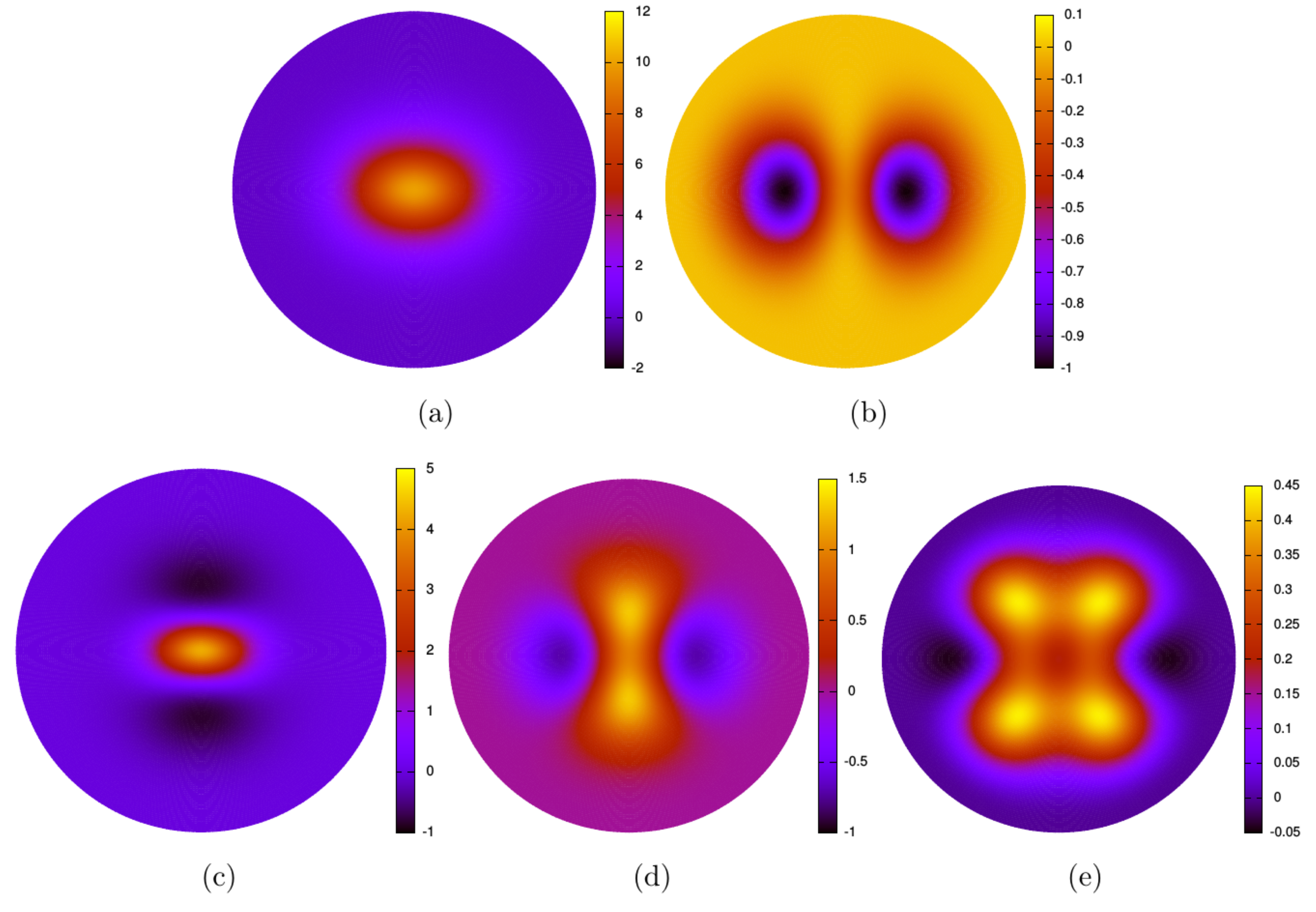}
   \caption{Component $\hat{h}_{xx}$ in the $x = 0$ plane of the conformal metric in the Andersson-Moncrief gauge for all five
families of geons studied in this paper, namely the $(l,m,n) = (2,2,0)$, $(l,m,n) = (4,4,0)$ on the top panel and families I, II and
III on the bottom panel. The masses of the solutions pictured are around $\sim 0.2 L$, and they were computed at our best
numerical resolutions.}
   \label{fig:gallery}
\end{figure}

\section{Conclusion}

In this paper, we presented both perturbative and numerical geons in asymptotically AdS spacetimes. We use perturbative approach
results at first order to seed an iterative solver of the Einstein-Andersson-Moncrief system, working in a
gauge combining maximal slicing and spatial harmonic coordinates. Monitoring precisely numerical errors, we were able to construct
geons with different levels of radial and angular excitations at unprecedentedly high amplitudes, reaching masses of order $\sim
1/2$ of the AdS radius. In particular, we gave an independent construction of the fully non-linear $(l,m,n) = (2,2,0)$ geons that
were constructed solely in \cite{Horowitz14}. Although we disagree in some ways with \cite{Horowitz14}, the excellent agreement
between our analytical and numerical procedures (see \Sref{res}) makes us very confident in the correctness of our results.

We also presented the so-called Andersson-Moncrief gauge and discussed its theoretical
motivations as well as its numerical implementation. The link between this gauge and the harmonic gauge
enforced with the popular De-Turck method is derived in appendix \ref{DeTurck}. Last but not least we extended the numerical
constructions of fully non-linear geons to the $(l,m,n) = (4,4,0)$ case, as well as to the three excited families exhibiting one
radial node. The literature about these solutions is quite controversial at the moment, since \cite{Dias16a} argues that they
cannot exist while \cite{Rostworowski16,Rostworowski17} supports the idea of their existence. We hope that our numerical
construction of these excited geons will put an end to the debate, as we did construct them in our fully non-linear numerical
simulations. All five families of geons we built numerically are pictured on \Fref{fig:gallery}.

A very interesting continuation of this work would be to use our helical stationary solutions as initial data for an evolution code in
AdS. Connection with \cite{Andersson03} would be straightforward as we already use 3+1 formalism in the Andersson-Moncrief gauge.
This is a daunting task though, but it would give a definitive answer to the problem of purely gravitational islands of stability,
and it would be very enlightening to see a high amplitude geon evolving periodically without ever collapsing to a black hole or
developing an instability. Our intuition is that geons will bring a whole lot of interesting features in the future investigations
on the AdS instability problem beyond spherical symmetry.

\ack

We would like to thank the referees for their useful and relevant remarks. We are very grateful to Eric Gourgoulhon, Alexandre Le
Tiec and Silvano Bonazzola for enlightening discussions. G. M. is indebted to Fabrice Roy and Marco Mancini for their numerical
support. G. F. thanks the kind hospitality of the LUTH research group at Paris Observatory in Meudon during his two-year
Marie-Curie fellowship. This research has been supported in part by OTKA Grant No. K 101709 and by the Marie Curie Actions Intra
European Fellowship of the European Community's Seventh Framework Programme under contract number PIEF-GA-2013-621992. This work
was granted access to the HPC resources of MesoPSL financed by the Region Ile de France. This work was granted access to the HPC
resources of CINES.

\appendix

\section{Connection between De Turck and Andersson-Moncrief gauges}
\label{DeTurck}

Instead of \Eref{K} and \Eref{V}, the harmonic gauge enforces
\begin{equation}
   \xi^\alpha \equiv g^{\mu\nu}(\Gamma_{\mu\nu}^\alpha - \bar{\Gamma}_{\mu\nu}^\alpha) = 0.
\end{equation}

We can 3+1 decompose this vector into $\xi^\alpha = \zeta u^\alpha + \chi^\alpha$ where $u^\alpha$ is the normal vector to
$\Sigma_t$ and $\chi^\alpha$ is the spatial projection of $\xi^\alpha$, i.e. $u_{\alpha}\chi^\alpha = 0$. We can then compute $\zeta$ with\footnote{Formulas
from Appendix B of \cite{Alcubierre08} might help.}
\begin{equation}
   \zeta = -u_\alpha \xi^\alpha = N\xi^t = -K -\frac{1}{N^2}\mathcal{L}_m N - \frac{2}{N}\beta^i\partial_i \ln \bar{N} .
\end{equation}
As for the spatial components of $\chi^\alpha$ it comes :
\begin{eqnarray}
\fl\nonumber   \chi^i &=& \xi^i - \zeta u^i = \xi^i + \beta^i \xi^t \\
\fl\nonumber   &=& V^i - \frac{1}{N^2}\mathcal{L}_m\beta^i - \partial^i \ln N + \frac{\beta^j\beta^k}{N^2}\bar{\Gamma}_{jk}^i + \frac{1}{N^2}(\bar{N}\bar{\gamma}^{ij}\partial_j \bar{N} + \bar{\beta}^j\bar{D}_j \bar{\beta}^i)\\
\fl&&- \frac{2\beta^j}{N^2}((\beta^j- \bar{\beta}^j)\partial_j\ln \bar{N} + \bar{D}_j \bar{\beta}^i ).
\end{eqnarray}
What is remarkable is that symbolically
\begin{equation}
   (\zeta,\chi^i) = (-K,V^i) + \tn{terms in } \partial N, \partial \beta^i,
   \label{remarkable}
\end{equation}
so the Andersson-Moncrief gauge catches essentially the 3+1 decomposition of the De Turck vector.

On the other hand, the De Turck method doesn't solve the Einstein equation but
\begin{equation}
   G_{\alpha\beta} - \nabla_{(\alpha}\xi_{\beta)} = 0.
\end{equation}
We then 3+1 decompose the second term into :
\begin{eqnarray}
   \gamma^\alpha_{\mu}\gamma^\beta_{\nu}\nabla_{(\alpha}\xi_{\beta)} &=& -\zeta K_{\mu\nu} + D_{(\mu}\chi_{\nu)},\\
   \gamma^\alpha_{\mu}u^\beta\nabla_{(\alpha}\xi_{\beta)} &=& -\frac{1}{2}D_{\mu}\zeta + \frac{1}{2}\zeta D_{\mu}\ln N + K_{\mu\nu}\chi^\nu + \frac{1}{2N}\mathcal{L}_m\chi_\mu, \\
   u^\alpha u^\beta\nabla_{(\alpha}\xi_{\beta)} &=& -\frac{1}{N}\mathcal{L}_m \zeta - \chi^\mu D_\mu \ln N,\\
   \nabla_{\alpha} \xi^\alpha &=& \frac{1}{N}\mathcal{L}_m \zeta - \zeta K + D_{\mu}\chi^\mu + \chi^\mu D_{\mu}\ln N,
\end{eqnarray}
which gives the 3+1 Einstein-De Turck system :
\begin{eqnarray}
   \label{hamDT}
\fl   R + K^2 - K_{ij}K^{ij} - 2\Lambda + \frac{1}{N}\mathcal{L}_m \zeta + \zeta K - D_i \chi^i + \chi^i D_i \ln N = 0,\\
\fl   D_j K^j_i - D_i K - \frac{1}{2}D_i \zeta + \frac{1}{2}\zeta D_i \ln N + K_{ij}\chi^j + \frac{1}{2N}\mathcal{L}_m \chi_i = 0, \\
\label{evoDT}
\fl   \mathcal{L}_m K_{ij} = -D_iD_jN + N( R_{ij} + K K_{ij} - 2 K_{ij}K^k_j - \Lambda \gamma_{ij} + \zeta K_{ij} - D_{(i}\chi_{j)}) = 0.
\end{eqnarray}
Comparing with our Einstein-Andersson-Moncrief system of equations \eref{ham2}-\eref{evo2}, in the light of \Eref{remarkable}, it is remarkable that in
the Hamiltonian constraint and evolution equation, the same terms in $K$ are suppressed while the same terms in $D_iV_j$ are
generated. This demonstrates the close relationship between Andersson-Moncrief and De Turck method.

\section{Regularization of first order gauge-fixing}
\label{gaugereg}

Undertaking the same reasoning as in \Sref{reg}, it happens that \Eref{eqK} behaves like $O(\Omega^{-1})$ at the AdS boundary. So
after multiplication by $\Omega$, it takes the regularized form :
\begin{eqnarray}
   \fl\nonumber   \Omega[2 \hat{N}^2 \hat{\gamma}^{ij} \partial_i\partial_j \alpha + 2\partial_i \hat{\beta}^i \hat{\beta}^j \partial_j \alpha + 2\partial_i (\hat{N}^2 \hat{\gamma}^{ij})\partial_j \alpha + 2 \partial_i \hat{\beta}^i + (\hat{\beta}^k + \{\hat{N}^2 \hat{\gamma}^{kl} + \hat{\beta}^k \hat{\beta}^l\}\partial_l\alpha)\hat{\gamma}^{ij}\partial_k \hat{\gamma}_{ij}]\\
   - 6 \Omega_i[\hat{\beta}^i + (\hat{N}^2 \hat{\gamma}^{ij} + \hat{\beta}^i \hat{\beta}^j)\partial_j \alpha] = 0.
      \label{eqKreg}
\end{eqnarray}
A suitable boundary condition is to require $\alpha$ to be zero  at the AdS boundary.

As for \Eref{eqV}, it becomes :
\begin{equation}
   \fl   - \Omega^2 \hat{\gamma}^{kl} \partial_k\partial_l \xi^i - \hat{\gamma}^{kl}(\Omega\mathcal{L}_{\xi}\hat{\Gamma}_{kl}^i - \hat{\Gamma}_{kl}^m \partial_m \xi^i)  - (\hat{\Gamma}_{kl}^i - \hat{\bar{\Gamma}}_{kl}^i)(\Omega\mathcal{L}_{\xi}\hat{\gamma}^{kl} + 2 \hat{\gamma}^{kl}\xi^m \partial_m\Omega) + \Omega \hat{V}^i = 0.
   \label{eqVreg}
\end{equation}
A suitable boundary condition is to require $\xi^i$ to be zero  at the AdS boundary.

\section{Regularization of $T^{CFT}_{\alpha\beta}$}
\label{tcftreg}

In order to compute numerically $T^{CFT}_{\alpha\beta}$ by \Eref{tcft}, we need to use only regular, non-diverging quantities.
First, let us introduce $r_{\alpha}$ the unit normal to hypersurfaces $r=cst$, its acceleration $a_{\alpha}$, $q_{\alpha\beta}$
the metric induced by $g_{\alpha\beta}$ and $\Theta_{\alpha\beta}$ the corresponding extrinsic curvature
\begin{eqnarray}
   r_{\alpha} = \frac{\partial_{\alpha}r}{\sqrt{g^{\mu\nu}\partial_{\mu}r \partial_\nu r}},\\
   a_{\alpha} = r^\mu \nabla_{\mu}r_{\alpha},\\
   q_{\alpha\beta} = g_{\alpha\beta} - r_{\alpha}r_{\beta},\\
   \Theta_{\alpha\beta} = -\nabla_{\beta}r_{\alpha} + a_{\alpha}r_\beta .
\end{eqnarray}
Corresponding regularized quantities are then
\begin{eqnarray}
   \hat{r}_{\alpha} &\equiv& \Omega r_{\alpha} = \frac{\partial_{\alpha}r}{\sqrt{\hat{g}^{\mu\nu}\partial_{\mu}r \partial_\nu r}},\\
   \hat{a}_{\alpha} &\equiv& \Omega a_{\alpha} = r^\mu [\partial_{\mu}(\Omega \hat{r}_{\alpha}) - \hat{\Gamma}_{\mu\alpha}^\nu \hat{r}_{\nu} - 2 \hat{r}_{\alpha}\Omega_\mu],\\
   \hat{q}_{\alpha\beta} &\equiv& \Omega^2 q_{\alpha\beta} = \hat{g}_{\alpha\beta} - \hat{r}_{\alpha}\hat{r}_{\beta},\\
   \hat{\Theta}_{\alpha\beta} &\equiv& \Omega^2 \Theta_{\alpha\beta} = -\partial_{\beta}(\Omega \hat{r}_{\alpha}) + \hat{\Gamma}_{\alpha\beta}^\mu \hat{r}_\mu + 2 \hat{r}_{\alpha} \Omega_\beta + \hat{a}_\alpha \hat{r}_\beta,
\end{eqnarray}
where $\hat{\Gamma}_{\alpha\beta}^\mu = \Omega \Gamma_{\alpha\beta}^\mu$.
The Riemann tensor of $g_{\alpha\beta}$ can be regularized as follows
\begin{equation}
\fl   \hat{R}_{\alpha\beta\mu\nu} \equiv \Omega^4 R_{\alpha\beta\mu\nu} = \hat{g}_{\alpha\rho} (\Omega\partial_\mu \hat{\Gamma}_{\beta\nu}^\rho - \Omega\partial_\nu \hat{\Gamma}_{\beta\mu}^\rho - \hat{\Gamma}_{\beta\nu}^\rho \Omega_\mu + \hat{\Gamma}_{\beta\mu}^\rho \Omega_\nu) + \hat{g}_{\alpha\rho}(\hat{\Gamma}_{\sigma\mu}^\rho \hat{\Gamma}_{\beta\nu}^\sigma - \hat{\Gamma}_{\sigma\nu}^\rho \hat{\Gamma}_{\beta\mu}^\sigma),
\end{equation}
which allows us to recover the regularized Riemann tensor of $q_{\alpha\beta}$ via the Gauss relation \cite{Gourgoulhon07}
\begin{equation}
   \hat{\mathcal{R}}_{\alpha\beta\gamma\delta} \equiv \Omega^4 \mathcal{R}_{\alpha\beta\gamma\delta} = \hat{q}^\mu_\alpha \hat{q}^\nu_\beta \hat{q}^\rho_\gamma \hat{q}^\sigma_\delta \hat{R}_{\mu\nu\rho\sigma} + \hat{\Theta}_{\alpha\gamma}\hat{\Theta}_{\beta\delta} - \hat{\Theta}_{\alpha\delta}\hat{\Theta}_{\beta\gamma}.
\end{equation}
From this, it is straightforward to get the regularized Eintein tensor of $q_{\alpha\beta}$, $\hat{\mathcal{G}}_{\alpha\beta} =
\Omega^2 \mathcal{G}_{\alpha\beta}$. The quasilocal stress energy is then given by
\begin{equation}
   8\pi G T^{CFT}_{\alpha\beta} = \frac{1}{\Omega^2}\left(\hat{\Theta}_{\alpha\beta} - \hat{\Theta}\hat{q}_{\alpha\beta} - \frac{2}{L}\hat{q}_{\alpha\beta} + L \hat{\mathcal{G}}_{\alpha\beta}\right),
\end{equation}
where $\hat{\Theta} = \hat{q}^{\alpha\beta}\hat{\Theta}_{\alpha\beta}$. Even if not obvious at first sight, the parenthesis is
$O(\Omega^3)$ near the AdS boundary. To compute this formula numerically, we first compute the parenthesis and check that it is
zero to machine precision at $r=L$. To compute the division by $\Omega^2$ (that vanishes at the boundary), we take advantage of
the spectral representation provided by the Kadath library and perform the division in coefficient space. In essence one uses the
non-local nature of the spectral representation. This operation brings some numerical errors that can be monitored (see
\Sref{res}).

\section{Mass and angular momentum tests of Kerr-AdS}
\label{kerr}

In this appendix, we test our numerical determination of both AMD and BK charges on the analytical Kerr-AdS metric. This allows us
to probe our absolute numerical precision on mass and angular momentum for a large number of different resolutions. The Kerr-AdS
metric expressed in conformal coordinates (\Eref{confcoor}) is :

\begin{eqnarray}
 ds^2 &=& -\frac{\Delta - (1-\rho^2)^4\Delta_{\theta}a^2\sin^2\theta}{(1-\rho^2)^2 \Sigma}dt^2 +
   \frac{4}{(1-\rho^2)^2}\frac{(1+\rho^2)^2\Sigma}{\Delta}dr^2\\
\nonumber   &+& \frac{\Sigma}{(1-\rho^2)^2\Delta_\theta}d\theta^2
   + \frac{\Delta_\theta(4r^2 + a^2(1-\rho^2)^2)^2 - \Delta a^2\sin^2\theta}{(1-\rho^2)^2 \Sigma\Xi^2}d\varphi^2 \\
\nonumber   &-& \frac{\Delta_\theta(1-\rho^2)^2(4r^2 + a^2(1-\rho^2)^2) - \Delta}{(1-\rho^2)^2\Sigma\Xi}2a\sin^2\theta dt d\varphi,
\end{eqnarray}
with
\begin{eqnarray}
   \Delta &=& (4r^2 + a^2(1-\rho^2)^2)(1+\rho^2)^2 - 4mr(1-\rho^2)^3,\\
   \Sigma &=& 4r^2 + (1-\rho^2)^2 a^2 \cos^2\theta,\\
   \Delta_\theta &=& 1 - \frac{a^2}{L^2}\cos^2\theta,\\
   \Xi &=& 1 - \frac{a^2}{L^2},
\end{eqnarray}
where we choose $m$ and $a$ to be positive without loss of generality.

As explained in \cite{Caldarelli00,Gibbons05}, the parameters $m$ and $a$ are not the mass $M$ and angular momentum $J$ of the black
hole, but are related to them via
\begin{equation}
   M = \frac{m}{\Xi^2} \quad \tn{and} \quad J = -\frac{am}{\Xi^2}.
\end{equation}
It is a strange but physical effect of Kerr-AdS : $J$ and $a$ have opposite sign\footnote{For geons, we also observed that $J$ and
$\Omega$ have opposite signs, however in the results presented in this paper we changed the sign of $J$, as it seems common in
the literature.}, because the frame dragging function $\omega = -g_{t\varphi}/g_{\varphi\varphi}$ is positive near the horizon but
negative and finite at the AdS boundary, whereas in asymptotically flat spacetimes, $\omega$ is positive everywhere and goes to
zero on the sphere at infinity.

Applying either the AMD or BK definitions for charge, a naive computation gives
\begin{equation}
   Q_{\partial_t}[\Sigma_t] = \frac{m}{\Xi} \quad \tn{and} \quad Q_{\partial_\varphi}[\Sigma_t] = J.
\end{equation}
The reason why $M \neq Q_{\partial_t}[\Sigma_t]$ is that the observer whose worldline is attached to $\partial_t$ at the AdS
boundary is not a zero angular-momentum observer (ZAMO). A boundary ZAMO has actually an angular velocity $\omega = -a/L^2$. This
spoils the charge computation, and the correct mass is given by the charge attached to the ZAMO worldline
\begin{equation}
   M = Q_{\partial_t + \omega\partial_\varphi}[\Sigma_t] = Q_{\partial_t}[\Sigma_t] + \omega J.
   \label{zamo}
\end{equation}

In order to test our charge computation numerically, we select a Kerr-AdS configuration with $m/L = 1$ and $a/L^2 = 0.5$, whose
analytical charges are $M^0/L = 16/9$ and $J^0/L^2 = -8/9$, and compute numerically both AMD and BK charges taking into
account \Eref{zamo}. For this configuration, the horizon lies at $\sim 0.37L$, so we use only one domain describing
$r\in[0.5,1]L$ to avoid the coordinate singularity.

\begin{figure}[t]
   \centering
   \includegraphics[width = \textwidth]{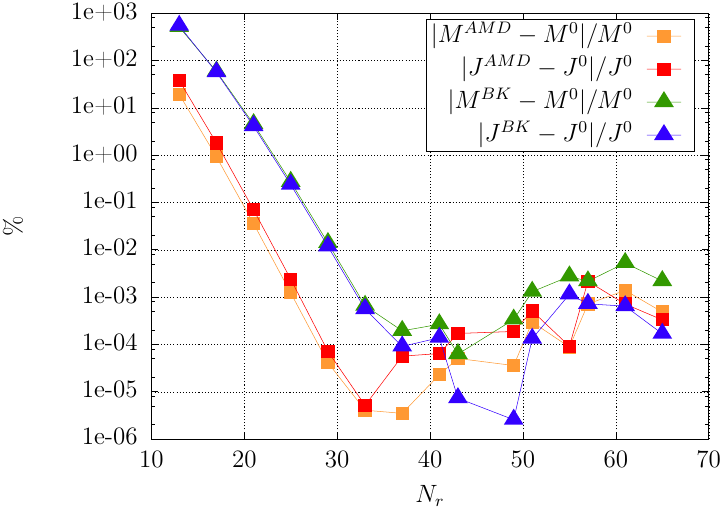}
   \caption{Relative differences between AMD, BK and analytical charges for the Kerr-AdS metric ($m/L = 1$, $a/L^2 = 1/2$) as a function
   of radial resolution. Angular resolution is fixed to $N_\theta = N_\varphi = 9$ assuming an octant symmetry.}
   \label{fig:kerr}
\end{figure}

In \Fref{fig:kerr}, we show the relative difference between analytical and numerical charges as a function of radial resolution.
As the results seemed quite insensitive to angular resolution, we fixed it at 9 points per octant. It is clear on this plot that both
AMD and BK charges converge exponentially to the analytical value up to $N_r = 37$-$41$, after which rounding errors start to
increase. At fixed resolution, BK is less precise than AMD, because of a more involved regularization procedure (see
\Sref{tcftreg}). Furthermore, our precision saturates at $\sim10^{-6}\%$, so that our absolute precision is around $\sim10^{-8}$ at a
resolution of 37x9x9. This is quite large for an analytical metric, but we can't do much better in double precision arithmetics,
since we need to perform several spectral operations like second order derivatives, divisions in coefficient space and surface
integration.

\newpage

\section*{References}

\bibliographystyle{unsrt}
\bibliography{biblio}

\end{document}